# Explainable Recommendation: A Survey and New Perspectives

**Yongfeng Zhang**
Rutgers University, USA
yongfeng.zhang@rutgers.edu

**Xu Chen**
Tsinghua University, China
xu-ch14@mails.tsinghua.edu.cn

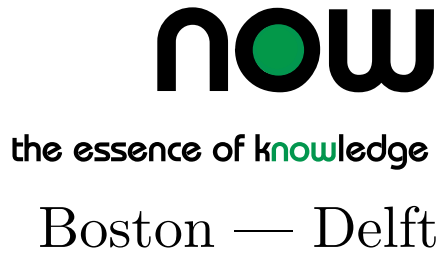

# Explainable Recommendation: A Survey and New Perspectives

Yongfeng Zhang[1] and Xu Chen[2]

[1] *Rutgers University, USA; yongfeng.zhang@rutgers.edu*
[2] *Tsinghua University, China; xu-ch14@mails.tsinghua.edu.cn*

---

ABSTRACT

Explainable recommendation attempts to develop models that generate not only high-quality recommendations but also intuitive explanations. The explanations may either be post-hoc or directly come from an explainable model (also called interpretable or transparent model in some contexts). Explainable recommendation tries to address the problem of *why*: by providing explanations to users or system designers, it helps humans to understand why certain items are recommended by the algorithm, where the human can either be users or system designers. Explainable recommendation helps to improve the transparency, persuasiveness, effectiveness, trustworthiness, and satisfaction of recommendation systems. It also facilitates system designers for better system debugging. In recent years, a large number of explainable recommendation approaches – especially model-based methods – have been proposed and applied in real-world systems.

In this survey, we provide a comprehensive review for the explainable recommendation research. We first highlight the position of explainable recommendation in recommender system research by categorizing recommendation problems into the 5W, i.e., what, when, who, where, and why. We then

---

conduct a comprehensive survey of explainable recommendation on three perspectives: 1) We provide a chronological research timeline of explainable recommendation, including user study approaches in the early years and more recent model-based approaches. 2) We provide a two-dimensional taxonomy to classify existing explainable recommendation research: one dimension is the information source (or display style) of the explanations, and the other dimension is the algorithmic mechanism to generate explainable recommendations. 3) We summarize how explainable recommendation applies to different recommendation tasks, such as product recommendation, social recommendation, and POI recommendation.

We also devote a section to discuss the explanation perspectives in broader IR and AI/ML research. We end the survey by discussing potential future directions to promote the explainable recommendation research area and beyond.

# 1

# Introduction

## 1.1 Explainable Recommendation

Explainable recommendation refers to personalized recommendation algorithms that address the problem of *why* – they not only provide users or system designers with recommendation results, but also explanations to clarify why such items are recommended. In this way, it helps to improve the transparency, persuasiveness, effectiveness, trustworthiness, and user satisfaction of the recommendation systems. It also facilitates system designers to diagnose, debug, and refine the recommendation algorithm.

To highlight the position of explainable recommendation in the recommender system research area, we classify personalized recommendation with a broad conceptual taxonomy. Specifically, personalized recommendation research can be classified into the 5W problems – when, where, who, what, and why, corresponding to time-aware recommendation (when), location-based recommendation (where), social recommendation (who), application-aware recommendation (what), and explainable recommendation (why), where explainable recommendation aims to answer *why*-type questions in recommender systems.

Explainable recommendation models can either be model-intrinsic or model-agnostic (Lipton, 2018; Molnar, 2019). The model-intrinsic approach develops interpretable models, whose decision mechanism is transparent, and thus, we can naturally provide explanations for the model decisions (Zhang *et al.*, 2014a). The model-agnostic approach (Wang *et al.*, 2018d), or sometimes called the post-hoc explanation approach (Peake and Wang, 2018), allows the decision mechanism to be a blackbox. Instead, it develops an explanation model to generate explanations after a decision has been made. The philosophy of these two approaches is deeply rooted in our understanding of human cognitive psychology – sometimes we make decisions by careful, rational reasoning and we can explain why we make certain decisions; other times we make decisions first and then find explanations for the decisions to support or justify ourselves (Lipton, 2018; Miller, 2019).

The scope of explainable recommendation not only includes developing transparent machine learning, information retrieval, or data mining models. It also includes developing effective methods to deliver the recommendations or explanations to users or system designers, because explainable recommendations naturally involve humans in the loop. Significant research efforts in user behavior analysis and human-computer interaction community aim to understand how users interact with explanations.

With this section, we will introduce not only the explainable recommendation problem, but also a big picture of the recommender system research area. It will help readers to understand what is unique about the explainable recommendation problem, what is the position of explainable recommendation in the research area, and why explainable recommendation is important to the area.

## 1.2 A Historical Overview

In this section, we will provide a historical overview of the explainable recommendation research. Though the term *explainable recommendation* was formally introduced in recent years (Zhang *et al.*, 2014a), the basic concept, however, dates back to some of the earliest works in personalized recommendation research. For example, Schafer *et al.* (1999) noted that recommendations could be explained by other items that the user is familiar with, such as *this product you are looking at is similar to these other products you liked before*, which leads to the fundamental idea of item-based collaborative filtering (CF); Herlocker *et al.* (2000) studied how to explain CF algorithms in MovieLens based on user surveys; and Sinha and Swearingen (2002) highlighted the role of transparency in recommender systems. Besides, even before explainable recommendation has attracted serious research attention, the industry has been using manual or semi-automatic explanations in practical systems, such as the *people also viewed* explanation in e-commerce systems (Tintarev and Masthoff, 2007a).

To help the readers understand the "pre-history" research of recommendation explanation and how explainable recommendation emerged as an essential research task in the recent years, we provide a historical overview of the research line in this section.

Early approaches to personalized recommender systems mostly focused on content-based or collaborative filtering (CF)-based recommendation (Ricci *et al.*, 2011). Content-based recommender systems model user and item profiles with various available content information, such as the price, color, brand of the goods in e-commerce, or the genre, director, duration of the movies in review systems (Balabanović and Shoham, 1997; Pazzani and Billsus, 2007). Because the item contents are easily understandable to users, it was usually intuitive to explain to users why an item is recommended. For example, one straightforward way is to let users know the content features he/she might be interested in the recommended item. Ferwerda *et al.* (2012) provided a comprehensive study of possible protocols to provide explanations for content-based recommendations.

However, collecting content information in different application domains is time-consuming. Collaborative filtering (CF)-based approaches (Ekstrand *et al.*, 2011), on the other hand, attempt to avoid this difficulty by leveraging *the wisdom of crowds*. One of the earliest CF algorithms is User-based CF for the GroupLens news recommendation system (Resnick *et al.*, 1994). User-based CF represents each user as a vector of ratings, and predicts the user's missing rating on a news message based on the weighted average of other users' ratings on the message. Symmetrically, Sarwar *et al.* (2001) introduced the Item-based CF method, and Linden *et al.* (2003) further described its application in Amazon product recommendation system. Item-based CF takes each item as a vector of ratings, and predicts the missing rating based on the weighted average of ratings from similar items.

Though the rating prediction mechanism would be relatively difficult to understand for average users, user- and item-based CF are somewhat explainable due to the philosophy of their algorithm design. For example, the items recommended by user-based CF can be explained as "users that are similar to you loved this item", while item-based CF can be explained as "the item is similar to your previously loved items". However, although the idea of CF has achieved significant improvement in recommendation accuracy, it is less intuitive to explain compared with content-based algorithms. Research pioneers in very early stages also noticed the importance of the problem (Herlocker and Konstan, 2000; Herlocker *et al.*, 2000; Sinha and Swearingen, 2002).

The idea of CF achieved further success when integrated with Latent Factor Models (LFM) introduced by Koren (2008) in the late 2000s. Among the many LFMs, Matrix Factorization (MF) and its variants were especially successful in rating prediction tasks (Koren *et al.*, 2009). Latent factor models have been leading the research and application of recommender systems for many years. However, though successful in recommendation performance, the "latent factors" in LFMs do not possess intuitive meanings, which makes it difficult to understand why an item got good predictions or why it got recommended out of other candidates. This lack of model explainability also makes it challenging to provide intuitive explanations to users, since it is hardly acceptable to tell users that we recommend an item only because it gets higher prediction scores by the model.

To make recommendation models better understandable, researchers have gradually turned to *Explainable Recommendation Systems*, where the recommendation algorithm not only outputs a recommendation list, but also explanations for the recommendations by working in an explainable way. For example, Zhang *et al.* (2014a) defined the *explainable recommendation* problem, and proposed an Explicit Factor

Model (EFM) by aligning the latent dimensions with explicit features for explainable recommendation. More approaches were also proposed to address the explainability problem, which we will introduce in detail in the survey. It is worthwhile noting that deep learning (DL) models for personalized recommendation have emerged in recent years. We acknowledge that whether DL models truly improve the recommendation performance is controversial (Dacrema *et al.*, 2019), but this problem is out of the scope of this survey. In this survey, we will focus on the problem that the black-box nature of deep models brings difficulty in model explainability. We will review the research efforts on explainable recommendation over deep models.

In a broader sense, the explainability of AI systems was already a core discussion in the 1980s era of "old" or logical AI research, when knowledge-based systems predicted (or diagnosed) well but could not explain why. For example, the work of Clancy showed that being able to explain predictions requires far more knowledge than just making correct predictions (Clancey, 1982). The recent boom in big data and computational power have brought AI performance to a new level, but researchers in the broader AI community have again realized the importance of *Explainable AI* in recent years (Gunning, 2017), which aims to address a wide range of AI explainability problems in deep learning, computer vision, autonomous driving systems, and natural language processing tasks. As an essential branch of AI research, this also highlights the importance of the IR/RecSys community to address the explainability issues of various search and recommendation systems. Moreover, explainable recommendation has also become a very suitable problem setting to develop new *Explainable Machine Learning* theories and algorithms.

## 1.3 Classification of the Methods

In this survey, we provide a classification taxonomy of existing explainable recommendation methods, which can help readers to understand the state-of-the-art of explainable recommendation research.

Specifically, we classify existing explainable recommendation research with two orthogonal dimensions: 1) The information source or display style of the explanations (e.g., textual sentence explanation, or visual explanation), which represents the human-computer interaction (HCI) perspective of explainable recommendation research, and 2) the model to generate such explanations, which represents the machine learning (ML) perspective of explainable recommendation research. Potential explainable models include the nearest-neighbor, matrix factorization, topic modeling, graph models, deep learning, knowledge reasoning, association rule mining, and others.

With this taxonomy, each combination of the two dimensions refers to a particular sub-direction of explainable recommendation research. We should note that there could exist conceptual differences between "how explanations are presented (display style)" and "the type of information used for explanations (information source)". In the context of explainable recommendation, however, these two principles are closely related to each other because the type of information usually determines how the explanations can be displayed. As a result, we merge these two principles into a single classification dimension. Note that among the possibly many classification taxonomies, this is just one that we think would be appropriate to organize the research on explainable recommendation, because it considers both HCI and ML perspectives of explainable recommendation research.

Table 1.1 shows how representative explainable recommendation research is classified into different categories. For example, the Explicit Factor Model (EFM) for explainable recommendation (Zhang *et al.*, 2014a) developed a matrix factorization method for explainable recommendation, which provides an explanation sentence for the recommended item. As a result, it falls into the category of "matrix factorization with textual explanation". The Interpretable Convolutional Neural Network approach (Seo *et al.*, 2017), on the other hand, develops a deep convolutional neural network model and displays item features to users as explanations, which falls into the category of "deep learning with user/item feature explanation". Another example is visually explainable recommendation (Chen *et al.*, 2019b), which proposes a deep model to generate image regional-of-interest explanations, and it belongs to the "deep learning with visual explanation" category. We also classify other research according to this taxonomy, so that readers can understand the relationship between existing explainable recommendation methods.

**Table 1.1:** A classification of existing explainable recommendation methods. The classification is based on two dimensions, i.e., the type of model for explainable recommendation (e.g., matrix factorization, topic modeling, deep learning, etc.) and the information/style of the generated explanation (e.g., textual sentence explanation, etc.). Note that due to the table space this is an incomplete enumeration of the existing explainable recommendation methods, and more methods are introduced in detail in the following parts of the survey. Besides, some of the table cells are empty because to the best of our knowledge there has not been a work falling into the corresponding combination

| | Methods for Explainable Recommendation | | | | | | | |
|---|---|---|---|---|---|---|---|---|
| Information/ style of the explanations | Neighbor-based | Matrix factorization | Topic modeling | Graph-based | Deep learning | Knowledge-based | Rule mining | Post-hoc |
| Relevant user or item | Herlocker *et al.*, 2000 | Abdollahi and Nasraoui, 2017 | | Heckel *et al.*, 2017 | Chen *et al.*, 2018c | Catherine *et al.*, 2017 | Peake and Wang 2018 | Cheng *et al.*, 2019a |
| User or item features | Vig *et al.*, 2009 | Zhang *et al.*, 2014a | McAuley and Leskovec, 2013 | He *et al.*, 2015 | Seo *et al.*, 2017 | Huang *et al.*, 2018 | Davidson *et al.*, 2010 | McInerney *et al.*, 2018 |
| Textual sentence explanation | | Zhang *et al.*, 2014a | | | Li *et al.*, 2017 | Ai *et al.*, 2018 | Balog *et al.*, 2019 | Wang *et al.*, 2018d |
| Visual explanation | | | | | Chen *et al.*, 2019b | | | |
| Social explanation | Sharma and Cosley, 2013 | | Ren *et al.*, 2017 | Park *et al.*, 2018 | | | | |
| Word cluster | | Zhang, 2015 | Wu and Ester 2015 | | | | | |

Due to the large body of related work, Table 1.1 is only an incomplete enumeration of explainable recommendation methods. For each "model – information" combination, we present one representative work in the corresponding table cell. However, in Sections 2 and 3 of the survey, we will introduce the details of many explainable recommendation methods.

## 1.4 Explainability and Effectiveness

Explainability and effectiveness could sometimes be conflicting goals in model design that we have to trade-off (Ricci *et al.*, 2011), i.e., we can either choose a simple model for better explainability, or choose a complex model for better accuracy while sacrificing the explainability. While recent evidence also suggests that these two goals may not necessarily conflict with each other when designing recommendation models (Bilgic *et al.*, 2004; Zhang *et al.*, 2014a). For example, state-of-the-art techniques – such as the deep representation learning approaches – can help us to design recommendation models that are both effective and explainable. Developing explainable deep models is also an attractive direction in the broader AI community, leading to progress not only in explainable recommendation research, but also in fundamental explainable machine learning problems.

When introducing each explainable recommendation model in the following sections, we will also discuss the relationship between explainability and effectiveness in personalized recommendations.

## 1.5 Explainability and Interpretability

Explainability and interpretability are closely related concepts in the literature. In general, interpretability is one of the approaches to achieve explainability. More specifically, Explainable AI (XAI) aims to develop models that can explain their (or other model's) decisions for system designers or normal users. To achieve the goal, the model can be either interpretable or non-interpretable. For example, interpretable models (such as interpretable machine learning) try to develop models whose decision mechanism is locally or globally transparent, and in this way, the model outputs are usually naturally explainable. Prominent examples of interpretable models include many linear models such as linear regression and tree-based models such as decision trees. Meanwhile, interpretability is not the only way to achieve explainability, e.g., some models can reveal their internal decision mechanism for explanation purpose with complex explanation techniques, such as neural attention mechanisms, natural language explanations, and many post-hoc explanation models, which are widely used in information retrieval, natural language processing, computer vision, graph analysis, and many other tasks. Researchers and practitioners may design and select appropriate explanation methods to achieve explainable AI for different tasks.

## 1.6 How to Read the Survey

Potential readers of the survey include both researchers and practitioners interested in explainable recommendation systems. Readers are encouraged to prepare with basic understandings of recommender systems, such as content-based recommendation (Pazzani and Billsus, 2007), collaborative filtering (Ekstrand *et al.*, 2011), and evaluation of recommender systems (Shani and Gunawardana, 2011). It is also beneficial to read other related surveys such as explanations in recommender systems from a user study perspective (Tintarev and Masthoff, 2007a), interpretable machine learning (Lipton, 2018; Molnar, 2019), as well as explainable AI in general (Gunning, 2017; Samek *et al.*, 2017).

The following part of the survey will be organized as follows. In Section 2 we will review explainable recommendation from a user-interaction perspective. Specifically, we will discuss different information sources that can facilitate explainable recommendation, and different display styles of recommendation explanation, which are closely related with the corresponding information source. Section 3 will focus on a machine learning perspective of explainable recommendation, which will introduce different types of models for explainable recommendation. Section 4 will introduce evaluation protocols for explainable recommendation, while Section 5 introduces how explainable recommendation methods are used in different real-world recommender system applications. In Section 6 we will summarize the survey with several important open problems and future directions of explainable recommendation research.

# 2

## Information Source for Explanations

In the previous section, we adopted a two-dimensional taxonomy to classify existing explainable recommendation research. In this section, we focus on the first dimension, i.e., the information source (or display style) of recommendation explanations. The second dimension (models/techniques for explainable recommendation) will be discussed in the next section.

An explanation is a piece of information displayed to users, explaining why a particular item is recommended. Recommendation explanations can be generated from different information sources and be presented in different display styles (Tintarev and Masthoff, 2015), e.g., a relevant user or item, a radar chart, a sentence, an image, or a set of reasoning rules. Besides, there could exist many different explanations for the same recommendation.

For example, Zhang *et al.* (2014a) generated (personalized) textual sentences as explanations to help users understand each recommendation; Wu and Ester (2015), Zhang (2015), and Al-Taie and Kadry (2014) provided topical word clouds to highlight the key features of a recommended item; Chen *et al.* (2019b) proposed visually explainable recommendations, where certain regions of a product image are highlighted as the visual explanations; Sharma and Cosley (2013) and Quijano-Sanchez *et al.* (2017) generated a list of social friends who also liked the recommended product as social explanations. In early research stages, Herlocker *et al.* (2000), Bilgic and Mooney (2005), and Tintarev and Masthoff (2007b) and McSherry (2005) adopted statistical histograms or pie charts as explanations to help users understand the rating distribution and the pros/cons of a recommendation. Du *et al.* (2019) provided a visual analytics approach to explainable recommendation for event sequences. Figure 2.1 shows several representative recommendation explanations.

In this section, we provide a summary of the different types of recommendation explanations to help readers understand what an explanation can look like in real-world settings. We also categorize the related work according to different explanation styles. More specifically, the following subsections present an overview of several frequently seen explanations in existing systems.

## 2.1 Relevant User or Item Explanation

We start from the very early stages of recommendation explanation research. In this section, we introduce explainable recommendation based on user- and item-based collaborative filtering (Cleger-Tamayo *et al.*, 2012; Resnick *et al.*, 1994; Sarwar *et al.*, 2001; Zanker and Ninaus, 2010) – two fundamental methods for personalized recommendation. Extensions of the two basic methods will also be introduced in this section.

User-based and item-based explanations are usually provided based on users' implicit or explicit feedback. In user-based collaborative filtering (Resnick *et al.*, 1994), we first find a set of similar users (i.e., neighbors) for the target user. Once the algorithm recommends an item to the target user, the explanation is that the user is similar to a group of "neighborhood" users, and these neighborhood users made good ratings on the recommended item. For example, Herlocker *et al.* (2000) compared the effectiveness of different display styles for explanations in user-based collaborative filtering. In this research, explanations can be displayed as an aggregated histogram of the neighbors' ratings, or be displayed as the detailed ratings of the neighbors, as shown in Figure 2.2. Recent model-based explainable recommendation approaches can generate more personalized and meticulously designed explanations than this, but this research illustrated the basic ideas of providing explanations in recommender systems.

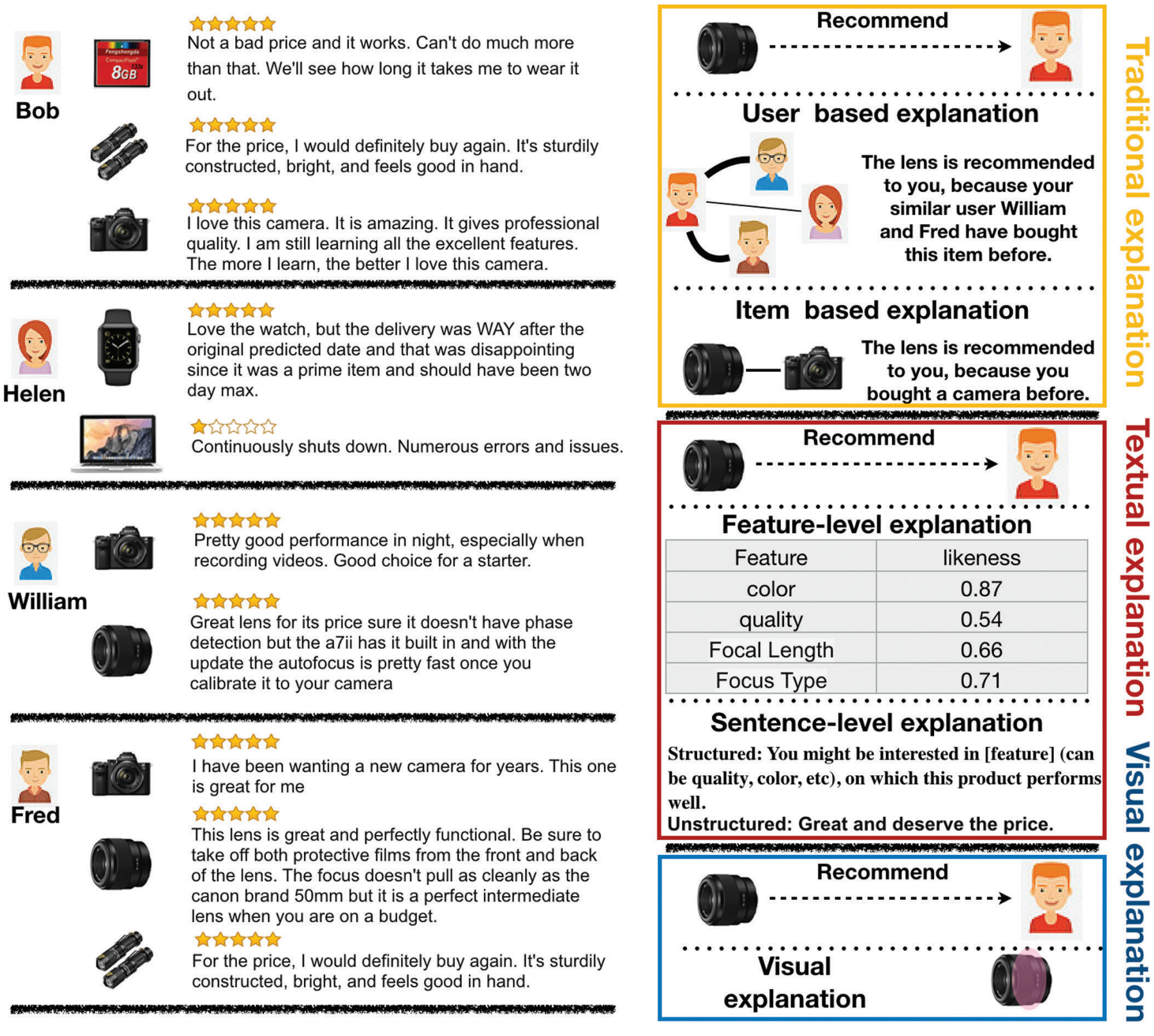

**Figure 2.1:** Different types of recommendation explanations. On the left panel, there are four example users, together with their purchased items and the corresponding reviews or ratings. On the right panel, we show some different display styles of the explanations, which are generated based on different information sources.

In item-based collaborative filtering (Sarwar *et al.*, 2001), explanations can be provided by telling the user that the recommended item is similar to some other items the user liked before, as shown in the left panel of Figure 2.3, where several highly rated movies by the user (4 or 5 stars) are shown as explanations. More intuitively, as in Figure 2.1, for the recommended item (i.e., the camera lens), a relevant-user explanation tells Bob that similar users William and Fred also bought this item, while a relevant-item explanation persuades Bob to buy the lens by showing the camera he already bought before.

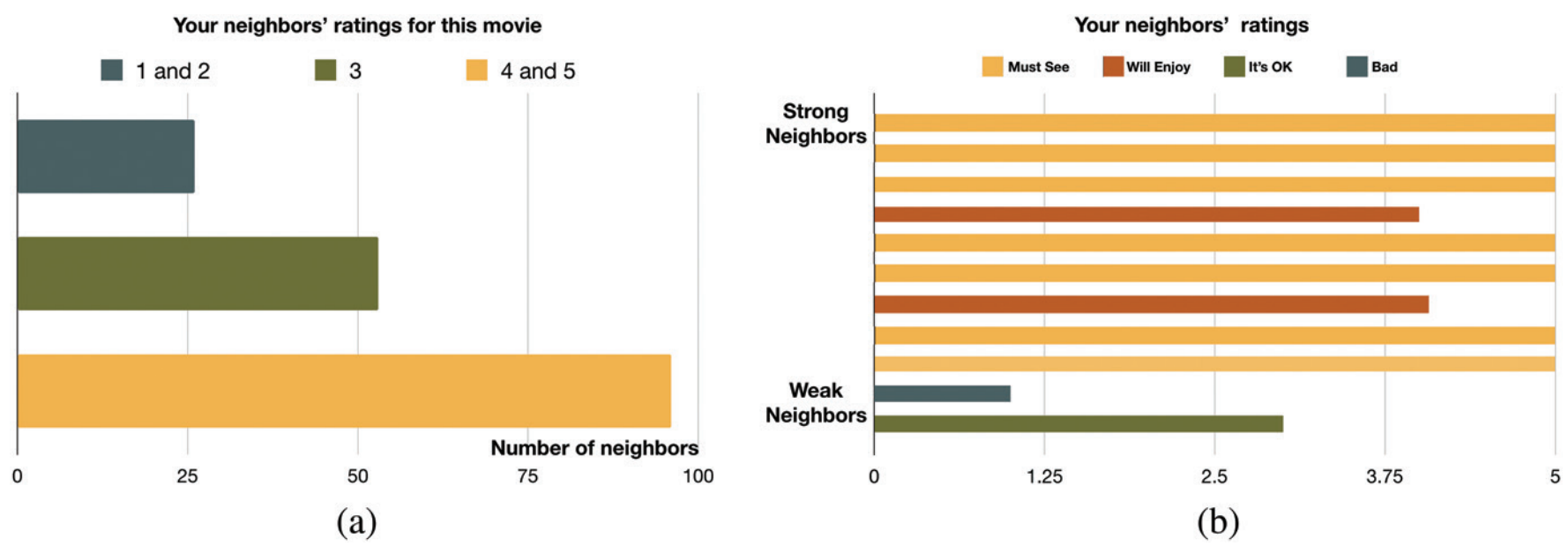

**Figure 2.2:** An example of explanations based on relevant users. (a) A histogram of the neighbors' ratings are displayed as an explanation for the recommended item, where the positive and negative ratings are correspondingly clustered, and the neutral ratings are displayed separately. Based on this explanation, a user can easily understand that the item is recommended because his/her neighbors made high ratings on the item. (b) An explanation for the movie "The Sixth Sense", where each bar represents the rating of a neighbor, and the $x$-axis represents a neighbor's similarity to the user. With this explanation, it would be easy to understand how a user's most similar neighbors rated the target movie (Herlocker *et al.*, 2000).

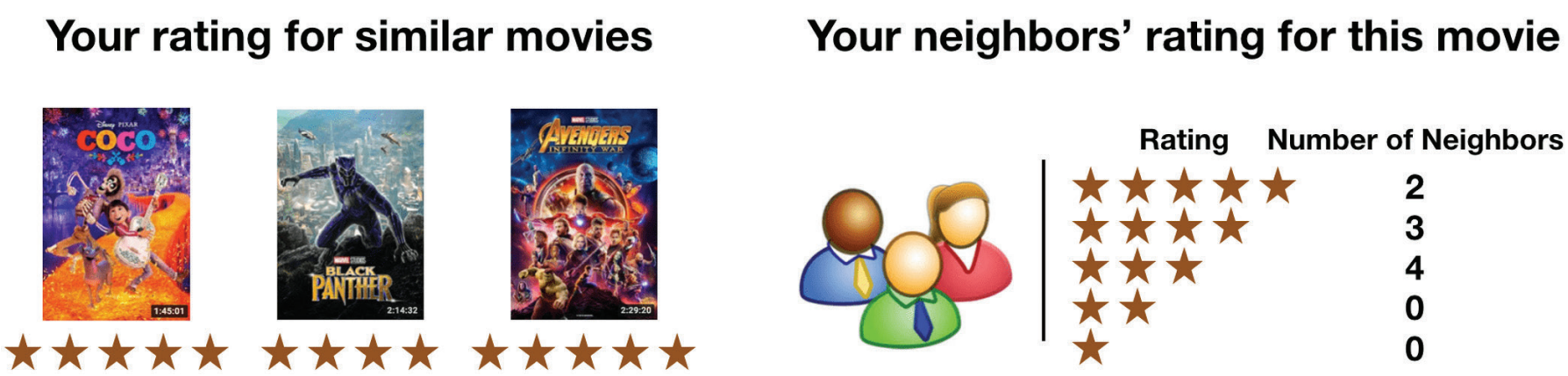

**Figure 2.3:** A comparison between relevant-item explanation (left) and relevant-user explanation (right) (Abdollahi and Nasraoui, 2017).

To study how explanations help in recommender systems, Tintarev (2007) developed a prototype system to study the effect of different types of explanations, especially the relevant-user and relevant-item explanations. In particular, the author proposed seven benefits of providing recommendation explanations, including transparency, scrutability, trustworthiness, effectiveness, persuasiveness, efficiency, and satisfaction.

Based on their user study, the author showed that providing appropriate explanations can indeed benefit the recommender system over these seven perspectives.

Usually, relevant-item explanations are more intuitive for users to understand because users are familiar with the items they interacted before. As a result, these items can serve as credible explanations for users. Relevant-user explanations, however, could be less convincing because the target user may know nothing about other "similar" users at all, which may decrease the trustworthiness of the explanations (Herlocker *et al.*, 2000). Besides, disclosing other users' information may also cause privacy problems in commercial systems. This problem drives relevant-user explanation into a new direction, which leverages social friend information to provide social explanations (Ren *et al.*, 2017; Tsai and Brusilovsky, 2018). For example, we can show a user with her friends' public interest as explanations for our social recommendations. In the following, we will review this research direction in the social explanation section (Section 2.6).

## 2.2 Feature-based Explanation

The feature-based explanation is closely related to content-based recommendation methods. In content-based recommendation, the system provides recommendations by matching the user profile with the content features of candidate items (Cramer *et al.*, 2008; Ferwerda *et al.*, 2012; Pazzani and Billsus, 2007). Content-based recommendations are usually intuitive to explain based on the features.

Depending on the application scenario, content-based recommendations can be generated from different item features. For example, movie recommendations can be generated based on movie genres, actors, or directors; while book recommendations can be provided based on book types, price, or authors. A conventional paradigm for feature-based explanation is to show users with the features that match the user's profile.

Vig *et al.* (2009) adopted movie tags as features to generate recommendations and explanations, as shown in Figure 2.4. To explain the recommended movie, the system displays the movie features and tells the user why each feature is relevant to her. The authors also designed a user study and showed that providing feature-based explanations can help to improve the effectiveness of recommendations. Furthermore, Ferwerda *et al.* (2012) conducted a user study, and the results supported the idea that explanations are highly correlated with user trust and satisfaction in the recommendations.

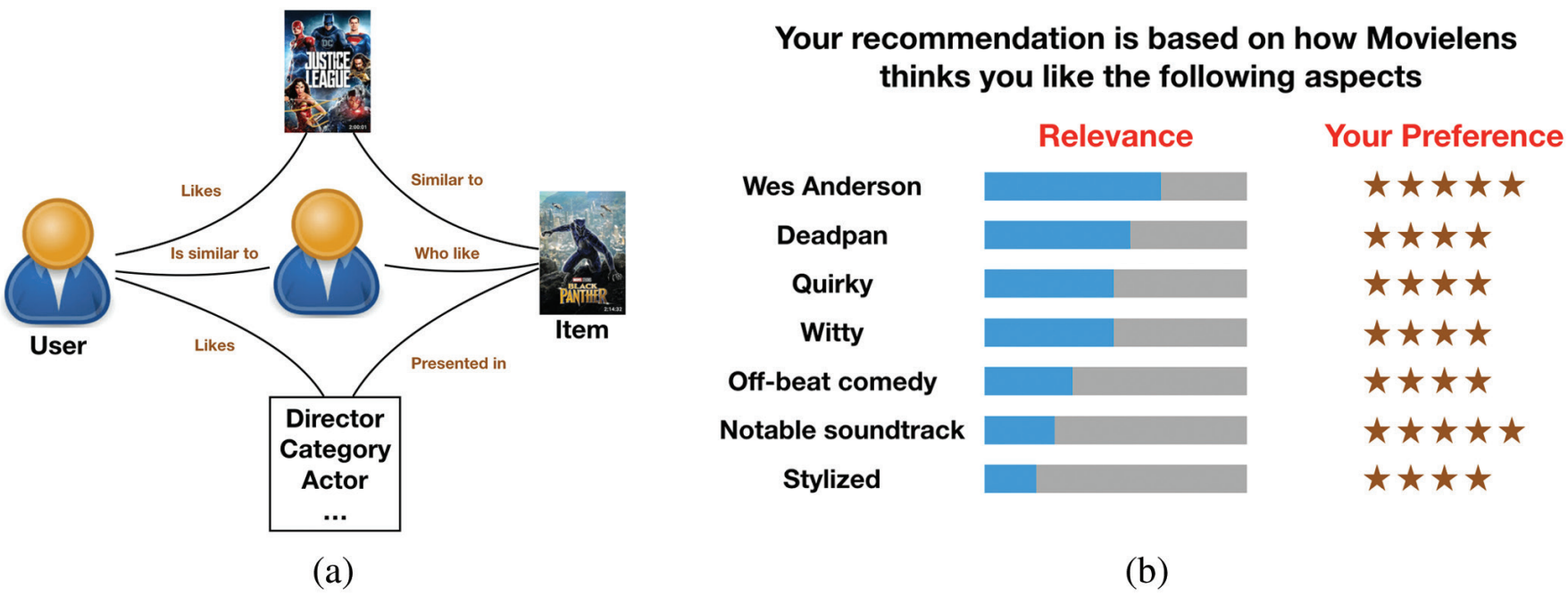

**Figure 2.4:** Tagsplanation: generating explanations based on content features such as tags (Vig *et al.*, 2009). (a) The basic idea of tags-based recommendation is to find the tags that a user likes, and then recommend items that match these tags. (b) The tagsplanations for a recommended movie *Rushmore*, where the relevant tags (features) are displayed as explanations.

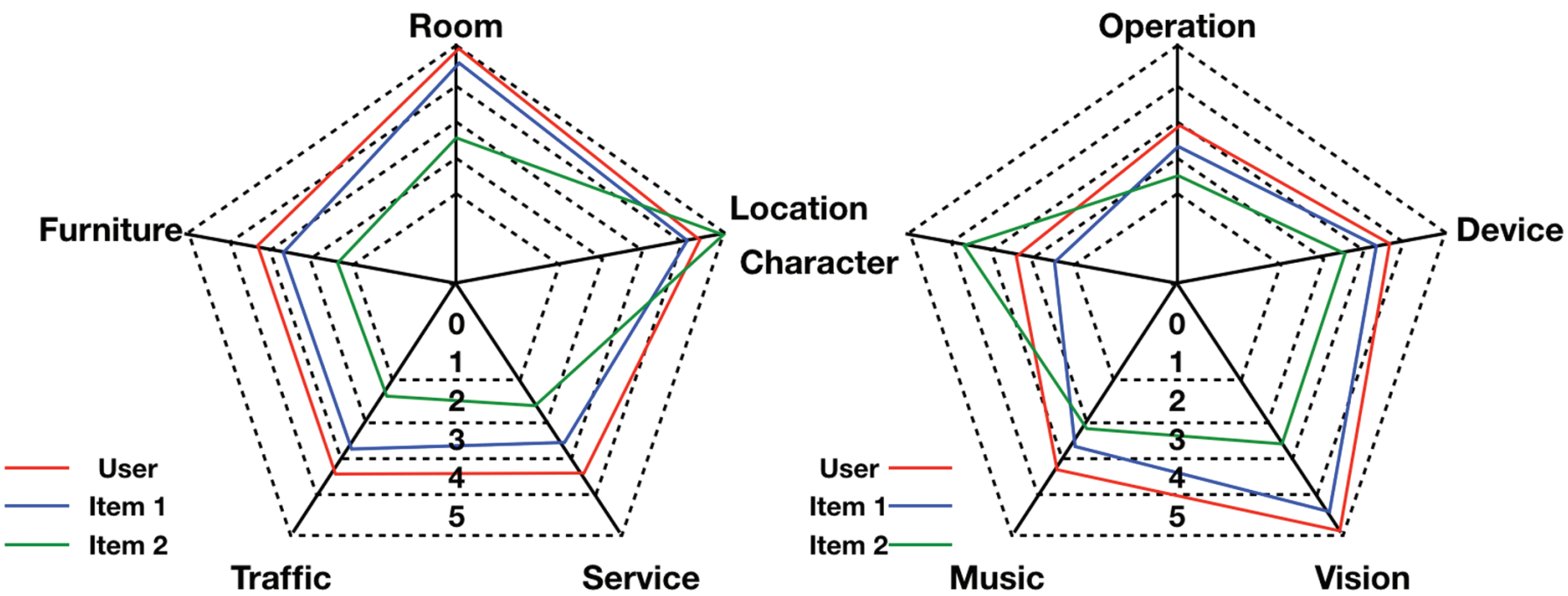

**Figure 2.5:** Using radar charts to explain a recommendation. The left figure shows hotel recommendations for a user, where item 1 is recommended because it satisfies the user preferences on almost all aspects. Similarly, the right figure shows video game recommendations, and also, item 1 is recommended (Hou *et al.*, 2018).

Content features can be displayed in many different explanation styles. For example, Hou *et al.* (2018) used radar charts to explain why an item is recommended and why others are not. As shown in Figure 2.5, a recommendation is explained in that most of its aspects satisfy the preference of the target user.

User demographic information describes the content features of users, and the demographic features can also be used to generate feature-based explanations. Demographic-based recommendation (Pazzani, 1999) is one of the earliest approaches to personalized recommendation systems. Recently, researchers have also integrated demographic methods into social media to provide product recommendations in social environments (Zhao *et al.*, 2014, 2016).

The demographic-based approach makes recommendations based on user demographic features such as age, gender, and residence location. Intuitively, an item recommended based on demographic information can be explained by the demographic feature(s) that triggered the recommendation, e.g., by telling the user that "80% of customers in your age bought this product". Zhao *et al.* (2014) represented products and users in the same demographic feature space, and used the weights of the features learned by a ranking function to explain the results; Zhao *et al.* (2016) further explored demographic information in social media environment for product recommendation with feature-based explanations.

## 2.3 Opinion-based Explanation

More and more user-generated contents have been accumulating on the Web, such as e-commerce reviews and social media posts, which help users to express their opinion on certain items or aspects. Researchers have shown that such information is quite beneficial in user profiling and recommendation (McAuley and Leskovec, 2013; Zheng *et al.*, 2017). Besides, it helps to generate finer-grained and more reliable explanations, which benefit users to make more informed decisions (Li *et al.*, 2017; Zhang *et al.*, 2014a). With this motivation, many models have been proposed to explain recommendations based on user-generated texts.

Methods in this direction can be broadly classified into aspect-level and sentence-level approaches, according to how the explanations are displayed. See Figure 2.1 for example, where aspect-level models present item aspects (such as color, quality) and their scores as explanations, while the sentence-level models directly present an explanation sentence to users about why the camera lens is recommended. We will focus on aspect-level explanation in this subsection, while sentence-level explanations will be introduced in the following subsection together with other natural language generation-based explanation models.

The aspect-level explanation is similar to feature-based explanation, except that aspects are usually not directly available in an item or user profile. Instead, they are extracted or learned as part of the recommendation model from – e.g., the reviews – and the aspects can be paired up with consumer opinions to express a clear sentiment on the aspect.

Researchers in the sentiment analysis community have explored both data mining and machine learning techniques for aspect-level sentiment analysis, which aims to extract aspect-sentiment pairs from text. For example, Hu and Liu (2004) proposed a frequent feature mining approach to aspect-level sentiment analysis, and Lu *et al.* (2011) proposed an optimization approach to construct context-aware sentiment lexicons automatically. A comprehensive review on sentiment analysis and opinion mining is summarized in Liu (2012). Based on these research efforts, Zhang *et al.* (2014b) developed a phrase-level sentiment analysis toolkit named *Sentires*[1] to extract "aspect–opinion–sentiment" triplets from reviews of a product domain. For example, given large-scale user reviews about mobile phones, the toolkit can extract triplets such as "noise–high–negative", "screen–clear–positive", and "battery_life–long–positive". The toolkit can also detect the contextual sentiment of the opinion words under different aspect words. For example, though "noise" paired with "high" usually represents a negative sentiment, when "quality" is paired with "high", however, it instead shows a positive sentiment. Based on the dictionary of aspect–opinion–sentiment triplets constructed by the program, it can further detect which triplets are contained in a review sentence.

Based on this toolkit, researchers developed different models for explainable recommendation. For example, Zhang *et al.* (2014a) proposed an explicit factor model for explainable recommendation, which presents aspect-opinion word clouds as explanations, such as "bathroom–clean", to highlight the performance of the recommended item on certain aspects. Wu and Ester (2015) developed a topic modeling approach for explainable hotel recommendations on TripAdvisor, which generates topical word cloud explanations on three hotel features (Location, Service, and Room), as shown in Figure 2.6. The word size in the word cloud is proportional to the sentiment opinion of the aspect. Ren *et al.* (2017) proposed a social collaborative viewpoint regression (sCVR) model for predicting item ratings based on user opinions and social relations, which provides viewpoints as explanations, where a viewpoint refers to a topic with a specific sentiment label. Combined with trusted user relations, it helps users to understand their friends' opinion about a particular item. Wang *et al.* (2018b) developed a multi-task learning solution for explainable recommendation, where two companion learning tasks of user preference modeling for recommendation and opinionated content modeling for explanation are integrated via a joint tensor factorization framework. We will introduce more about the model details in the explainable recommendation model section (Section 3).

[1] http://yongfeng.me/software/

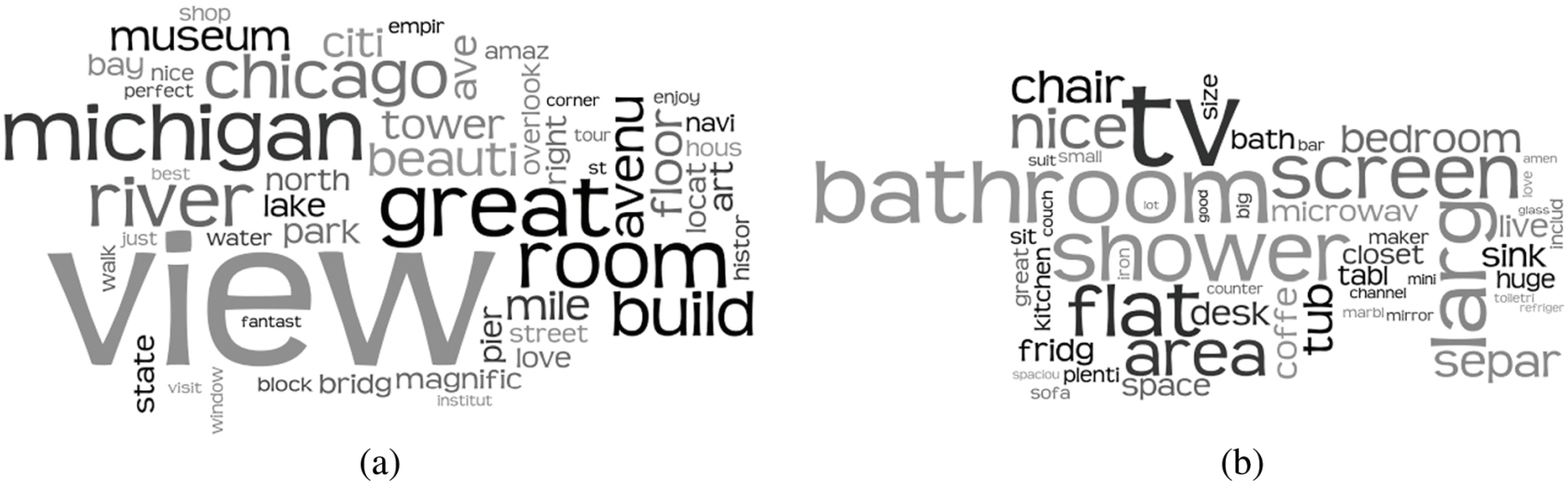

**Figure 2.6:** Word cloud explanation for hotel recommendations generated based on latent topic modeling with textual reviews. (a) Word cloud about the *Location* of the recommended hotel, and (b) word cloud about the *Service* of the recommended hotel. Courtesy image from Wu and Ester (2015).

## 2.4 Sentence Explanation

Sentence-level approach provides explanation sentences to users. This approach can be further classified into template-based approach and generation-based approach.

```
You might be interested in [feature],
on which this product performs well.
```

```
You might be interested in [feature],
on which this product performs poorly.
```

**Figure 2.7:** Generating sentence explanations with template-based methods.

Template-based approach first defines some explanation sentence templates, and then fills the templates with different words to personalize them for different users. For example, Zhang *et al.* (2014a) constructed explanations by telling the user that *You might be interested in* `feature`*, on which this product performs well.* In this template, the `feature` will be selected based on personalization algorithms to construct a personalized explanation, as shown in Figure 2.7. Based on the templates, the model can also provide "dis-recommendations" to let the user know why an item is not a good fit, by telling the user *You might be interested in* `feature`*, on which this product performs poorly.* Based on user studies, it shows that providing both recommendation and dis-recommendation explanations improve the persuasiveness and conversion rate of recommender systems.

Wang *et al.* (2018b) provided template-based explanations based on both feature and opinion words, for example, an explanation for Yelp restaurant recommendation could be *Its* `decor` *is* `[neat] [good] [nice]`*. Its* `sandwich` *is* `[grilled] [cajun] [vegan]`*. Its* `sauce` *is* `[good] [green] [sweet]`, where words in brackets are opinion words selected by the model to describe the corresponding item feature.

Tao *et al.* (2019a) further integrated regression trees to guide the learning of latent factor models, and used the learnt tree structure to explain the recommendations. They added predefined modifiers in front of the selected features to construct template-based explanations, such as *We recommend this item to you because its* `[good/excellent] [feature]` *matches with your* `[emphasize/taste]` *on* `[feature]`*.*

Gao *et al.* (2019) proposed an explainable deep multi-view learning framework to model multi-level features for explanation. They also adopted feature-based templates to provide explanations, and the features are organized in an industry-level hierarchy named Microsoft

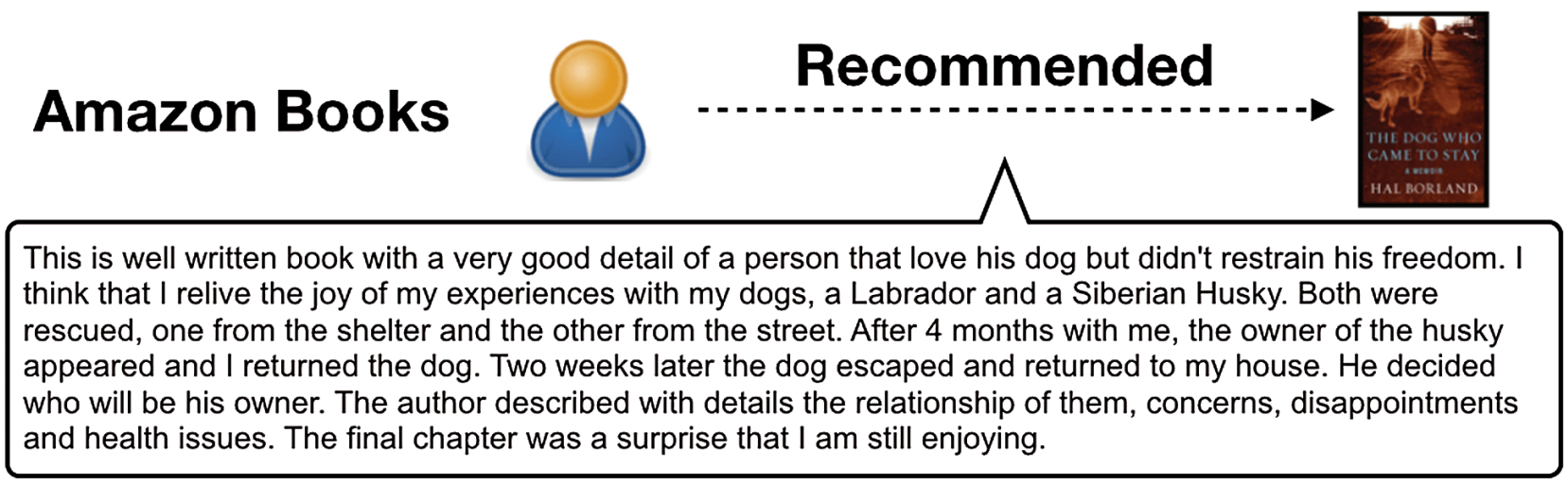

**Figure 2.8:** Generating sentence explanations based on natural language generation models such as LSTM (Costa *et al.*, 2018; Li *et al.*, 2017).

Concept Graph. The model improves accuracy and explainability simultaneously, and is capable of providing highly usable explanations in commercial systems. Technical details of the above models will be introduced in the explainable recommendation model section.

Based on natural language generation techniques, we can also generate explanation sentences directly without using templates. For example, Costa *et al.* (2018) generated explanation sentences based on long-short term memory (LSTM). By training over large-scale user reviews, the model can generate reasonable review sentences as explanations, as shown in Figure 2.8. Inspired by how people explain word-of-mouth recommendations, Chang *et al.* (2016) proposed a process to combine crowdsourcing and computation to generate personalized natural language explanations. The authors also evaluated the generated explanations in terms of efficiency, effectiveness, trust, and satisfaction. Li *et al.* (2017) leveraged gated recurrent units (GRU) to generate tips for a recommended restaurant in Yelp. According to the predicted ratings, the model can control the sentiment attitude of the generated tips, which help users understand the key features of the recommended items. Lu *et al.* (2018b) proposed a multi-task recommendation model, which jointly learns to perform rating prediction and recommendation explanation. The explanation module employs an adversarial sequence to sequence learning technique to encode, generate, and discriminate the user and item reviews. Once trained, the generator can be used to generate explanation sentences.

A lot of explanation generation approaches rely on user reviews as the training corpus to train an explanation generation model. However, user reviews are noisy, because not all of the sentences in a review are explanations or justifications of the users' decision-making process. Motivated by this problem, Chen *et al.* (2019a) proposed a hierarchical sequence-to-sequence model (HSS) for personalized explanation generation, which includes an auto-denoising mechanism that selects sentences containing item features for model training. Ni *et al.* (2019) introduced new datasets and methods to address this recommendation justification task. In terms of data, the authors proposed an extractive approach to identify review segments that justify users' intentions. In terms of generation, the authors proposed a reference-based Seq2Seq model with aspect-planning to generate explanations covering different aspects. The authors also proposed an aspect-conditional masked language model to generate diverse justifications based on templates extracted from justification histories.

## 2.5 Visual Explanation

To leverage the intuition of visual images, researchers have tried to utilize item images for explainable recommendation. In Figure 2.1, for example, to explain to Bob that the lens is recommended because of its collar appearance, the system highlights the image region corresponding to the necklet of the lens.

Lin *et al.* (2019) studied the explainable outfit recommendation problem, for example, given a top, how to recommend a list of bottoms (e.g., trousers or skirts) that best match the top from a candidate collection, and meanwhile generate explanations for each recommendation. Technically, this work proposed a convolutional neural network with a mutual attention mechanism to extract visual features of the outfits, and the visual features are fed into a neural prediction network to predict the rating scores for the recommendations. During the prediction procedure, the attention mechanism will learn the importance of different image regions, which tell us which regions of the image are taking effect when generating the recommendations, as shown in Figure 2.9(a).

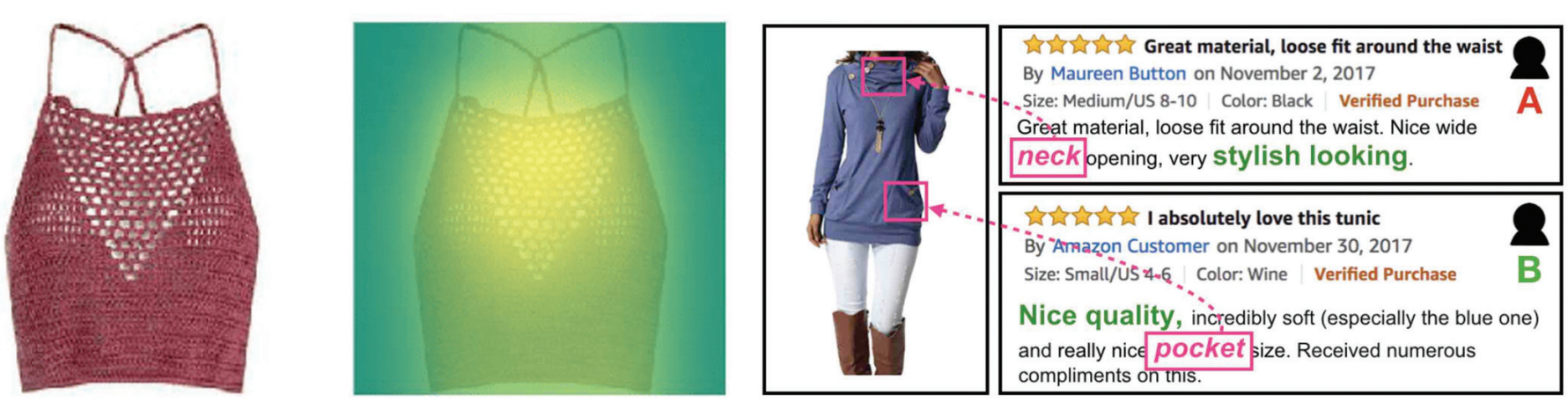

(a) Explainable outfit recommendation (b) Visually explainable recommendation

**Figure 2.9:** (a) Explainable outfit recommendation by Lin *et al.* (2019). The authors adopted a convolutional neural network with mutual attention to learn the importance of each pixel, and the importance scores tell us which pixels of the image take effect when generating the recommendations. (b) User interest on images are personalized: different users may care about different regions each for the sam image.

| Target Item | Textual Review | Visual Explanation | |
|---|---|---|---|
| | | VECF(-rev) | VECF |
| | I loved about the previous generation and ***expanded the toe box a little to improve the fit.*** great buy, highly recommended. | y, x, [x=13, y=13] | y, x, [x=9, y=1] |
| | ***They fit my stubby fingered hand pretty well.*** I bought the large and my hand measured 9.25&34 at the knuckles. | y, x, [x=10, y=5] | y, x, [x=3, y=1] |

**Figure 2.10:** Examples of the visual explanations in Chen *et al.* (2019b). Each row represents a recommended item. The first column lists the image of the item, and the second column shows the user's review on the item. The last two columns compare the region-of-interest explanations provided by two visually explainable recommendation models. In the review column, the bolded italic text shows how the review corresponds to the visual explanation.

Chen *et al.* (2019b) proposed *visually explainable recommendation* based on personalized region-of-interest highlights, as shown in Figure 2.10. The basic idea is that different regions of the product image may attract different users. As shown by the example in Figure 2.9(b), even for the same shirt, some users may care about the collar design while others may pay attention to the pocket. As a result, the authors adopted a neural attention mechanism integrated with both image and review information to learn the importance of each region of an image. The important image regions are highlighted in a personalized way as visual explanations for users.

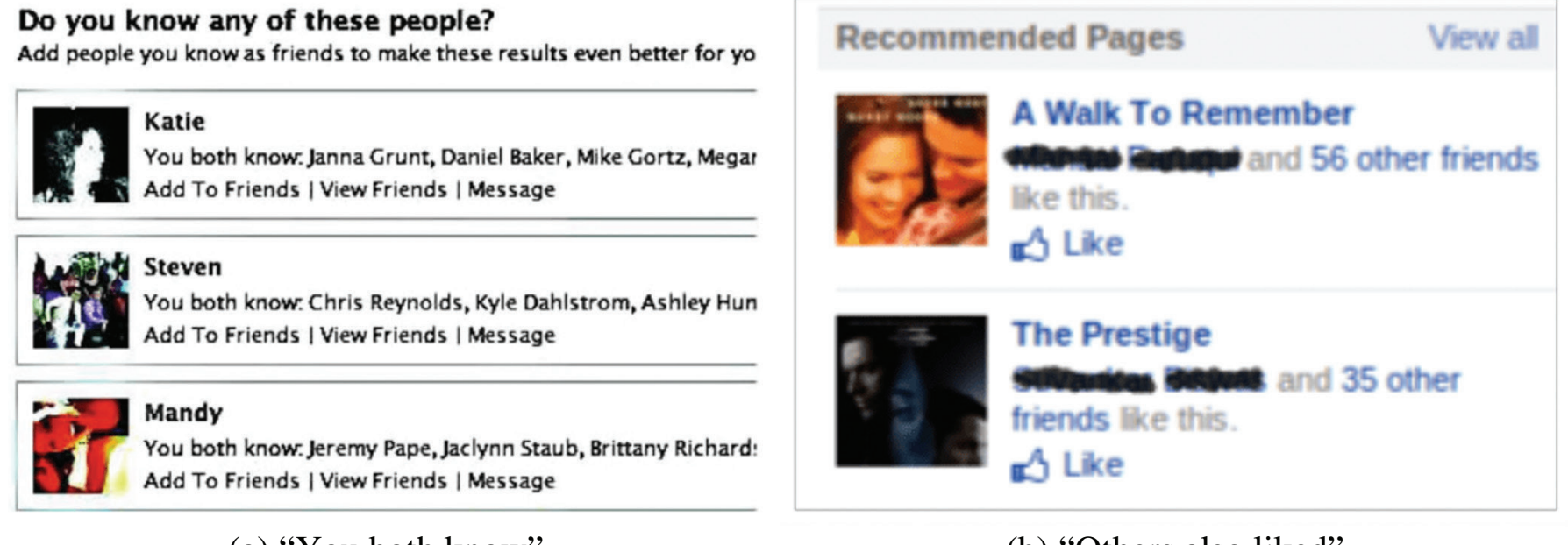

(a) "You both know" (b) "Others also liked"

**Figure 2.11:** Social explanations on Facebook. (a) Facebook provides friends in common as explanations when recommending a new friend to a user (Papadimitriou *et al.*, 2012). (b) Showing friends who liked the same item when recommending items to a user (Sharma and Cosley, 2013).

In general, the research on visually explainable recommendation is still at its initial stage. With the continuous advancement of deep image processing techniques, we expect that images will be better integrated into recommender systems for both better performance and explainability.

## 2.6 Social Explanation

As discussed in the previous subsections, a problem of relevant-user explanations is trustworthiness and privacy concerns, because the target user may have no idea about other users who have "similar interests". Usually, it will be more acceptable if we tell the user that his/her friends have similar interests in the recommended item. As a result, researchers proposed to generate social explanations, that is, explanations with the help of social information.

Papadimitriou *et al.* (2012) studied human-style, item-style, feature-style, and hybrid-style explanations in social recommender systems; they also studied geo-social explanations to combine geographical data with social data. For example, Facebook provides friends in common as explanations when recommending a new friend to a user (see Figure 2.11(b)). Sharma and Cosley (2013) studied the effects of social explanations in music recommendation by providing the target user with the number of friends that liked the recommended item (see Figure 2.11(b)). The authors found that explanations influence the likelihood of users checking out the recommended artists, but there is little correlation between the likelihood and the actual rating for the artist. Chaney *et al.* (2015) presented social Poisson factorization, a Bayesian model that incorporates a user's preference with her friends' latent influence, which provides explainable serendipity to users, i.e., pleasant surprise due to novelty.

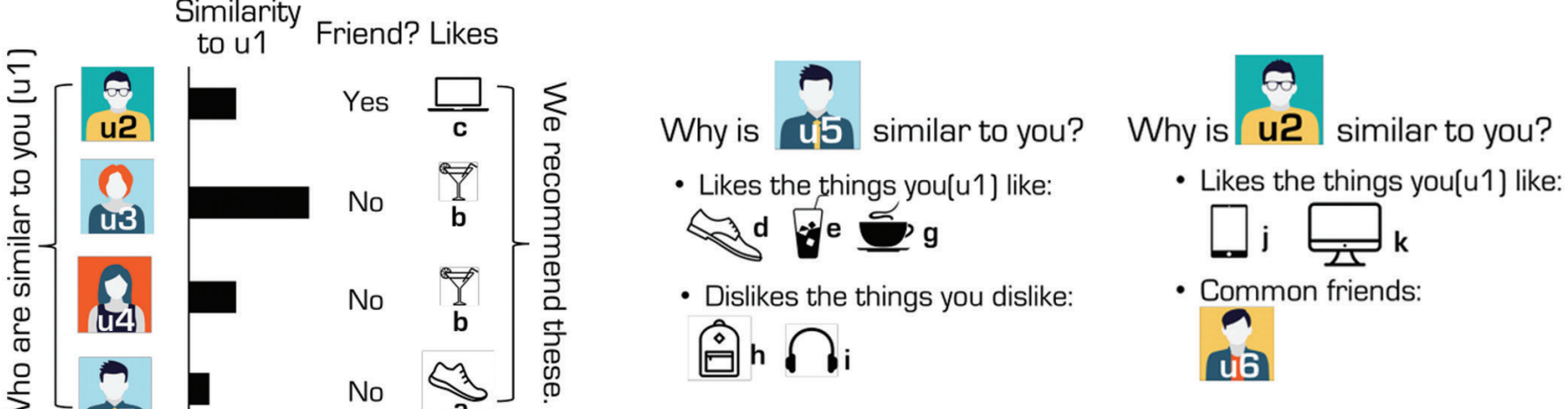

**Figure 2.12:** Explanations based on similar users, where similar users can be social friends or users that have the same preference on the same subset of products (Park *et al.*, 2018).

Except for friend recommendation, social explanations also take effect in other social network scenarios. For example, Park *et al.* (2018) proposed a unified graph structure to exploit both rating and social information to generate explainable product recommendations. In this framework, a recommendation can be explained based on the target user's friends who have similar preferences, as shown in Figure 2.12. Quijano-Sanchez *et al.* (2017) introduced a social explanation system applied to group recommendation, which significantly increased the likelihood of the user acceptance, the user satisfaction, and the system efficiency to help users make decisions. Wang *et al.* (2014) generates social explanations such as "*A and B also like the item*". They proposed to find an optimal set of users as the most persuasive social explanation. Specifically, a two-phase ranking algorithm is proposed, where the first phase predicts the persuasiveness score of a single candidate user, and the second phrase predicts the persuasiveness score of a set of users based on the predicted persuasiveness of the individual users, by taking the marginal net utility of persuasiveness, credibility of the explanation and reading cost into consideration.

## 2.7 Summary

In this section, we introduced several styles of recommendation explanations, including: 1) Explanations based on relevant users or items, which present nearest-neighbor users or items as an explanation. They are closely related to the critical idea behind user-based or item-based collaborative filtering methods. 2) Feature-based explanation, which provides users with the item features that match the target user's interest profile. This approach is closely related to content-based recommendation methods. 3) Opinion-based explanation, which aggregates users' collective opinions in user generated contents as explanations. 4) Textual sentence explanation, which provides the target users with explanation sentences. The sentence could be constructed based on pre-defined templates or directly generated based on natural language generation models. 5) Visual explanations, which provide users with image-based explanations. The visual explanation could be a whole image or a region-of-interest highlight in the image. 6) Social explanations, which provide explanations based on the target user's social relations. They help to improve user trust in recommendations and explanations.

It should be noted that while 1 to 3 are usually bonded with certain types of recommendation algorithms, 4 to 6 focus more on how the explanations are shown to users, which are not necessarily generated by one particular type of algorithm. In the following section, we will introduce more details of different explainable recommendation models.

# 3

# Explainable Recommendation Models

In the previous section, we have seen different types of explanations in recommender systems. Our next step is to describe how such explanations can be generated.

Explainable recommendation research can consider the explainability of either the recommendation methods or the recommendation results. When considering the explainability of methods, explainable recommendation aims to devise interpretable models for increased transparency, and such models usually directly lead to the explainability of results. In this section, many of the factorization-based, topic modeling, graph-based, deep learning, knowledge-based, and rule mining approaches adopt this research philosophy – they aim to design intrinsic explainable models and understand how the recommendation process works.

Another philosophy for explainable recommendation research is that we only focus on the explainability of the recommendation results. In this way, we treat the recommendation model as a blackbox, and develop separate models to explain the recommendation results produced by this blackbox. In this section, the post-hoc/model-agnostic approach adopts this philosophy.

In the following, we first provide a very brief overview of machine learning for general recommender systems, so as to provide readers with necessary background knowledge. Then we provide a comprehensive survey on explainable recommendation models.

## 3.1 Overview of Machine Learning for Recommendation

Recent research on model-based explainable recommendation is closely related to machine learning for recommender systems. We first provide a brief overview of machine learning for personalized recommendations in this section. One of the classical ML models for recommendation is Latent Factor Model (LFM), based on Matrix Factorization (MF). It learns latent factors to predict the missing ratings in a user-item rating matrix. Representative matrix factorization methods include Singular Value Decomposition (SVD) (Koren, 2008; Koren *et al.*, 2009; Srebro and Jaakkola, 2003), Non-negative Matrix Factorization (NMF) (Lee and Seung, 1999, 2001), Max-Margin Matrix Factorization (MMMF) (Rennie and Srebro, 2005; Srebro *et al.*, 2005), Probabilistic Matrix Factorization (PMF) (Mnih and Salakhutdinov, 2008; Salakhutdinov and Mnih, 2008), and Localized Matrix Factorization (LMF) (Zhang *et al.*, 2013a,b). Matrix factorization methods are also commonly referred to as point-wise prediction, and they are frequently used to predict user explicit feedbacks, such as numerical ratings in e-commerce or movie review websites.

Pair-wise learning to rank is frequently used to learn the correct item rankings based on implicit feedback. For example, Rendle *et al.* (2009) proposed Bayesian Personalized Ranking (BPR) to learn the relative ranking of the purchased items (positive item) against unpurchased items (negative items). Rendle and Schmidt-Thieme (2010) further extended the idea to tensor factorization to model pairwise interactions. Except for pair-wise learning to rank, Shi *et al.* (2010) adopted list-wise learning to rank for collaborative filtering.

Deep learning and representation learning also gained much attention in recommendation research. For example, Cheng *et al.* (2016) proposed Wide and Deep Network, which combines shallow regression and multiple layer neural network for recommender systems. Deep Neural Network is also applied in real-world commercial systems such as the YouTube recommender system (Covington *et al.*, 2016). Besides, researchers also explored various deep architectures and information modalities for recommendation, such as convolutional neural networks over text (Zheng *et al.*, 2017) or images (Chen *et al.*, 2019b), recurrent neural networks over user behavior sequence (Hidasi *et al.*, 2016), auto-encoders (Wu *et al.*, 2016), and memory networks (Chen *et al.*, 2018c).

Though many publications are generated in this direction due to the recent popularity of deep learning, we acknowledge that it is controversial whether neural models really make progress in recommender system research (Dacrema *et al.*, 2019). Deep models work when sufficient data is available for training, such as side information or knowledge graphs in some research datasets or in the industry environment (Zhang *et al.*, 2019). Nevertheless, recommendation performance is not a key focus of this survey. Instead, we will focus on the explainability perspective of deep models. A lot of deep models have been developed for explainable recommendation, which we will introduce in the following subsections.

## 3.2 Factorization Models for Explainable Recommendation

In this section, we introduce how matrix/tensor-factorization and factorization machines are used for explainable recommendation.

A lot of explainable recommendation models have been proposed based on matrix factorization methods. One common problem of matrix factorization methods – or more generally, latent factor models – is that the user/item embedding dimensions are latent. Usually, we assume that the user and item representation vectors are embedded in a low-dimensional space, where each dimension represents a particular factor that affects user decisions. However, we do not explicitly know the exact meaning of each factor, which makes the predictions or recommendations challenging to explain.

To solve the problem, Zhang *et al.* (2014a) proposed Explicit Factor Models (EFM). The basic idea is to recommend products that perform well on the user's favorite features, as shown in Figure 3.1. Specifically, the proposed approach extracts explicit product features from user reviews, and then aligns each latent dimension of matrix factorization with an explicit feature, so that the factorization and prediction procedures are trackable to provide explicit explanations. The proposed approach can provide personalized recommendation explanations based on the explicit features, e.g., "The product is recommended because you are interested in a particular feature, and this product performs well on the feature". The model can even provide dis-recommendations by telling the user that "The product does not perform very well on a feature that you care about", which helps to improve the trustworthiness of recommendation systems. Because user preferences on item features are dynamic and may change over time, Zhang *et al.* (2015b) extended the idea by modeling the user's favorite features dynamically on daily resolution.

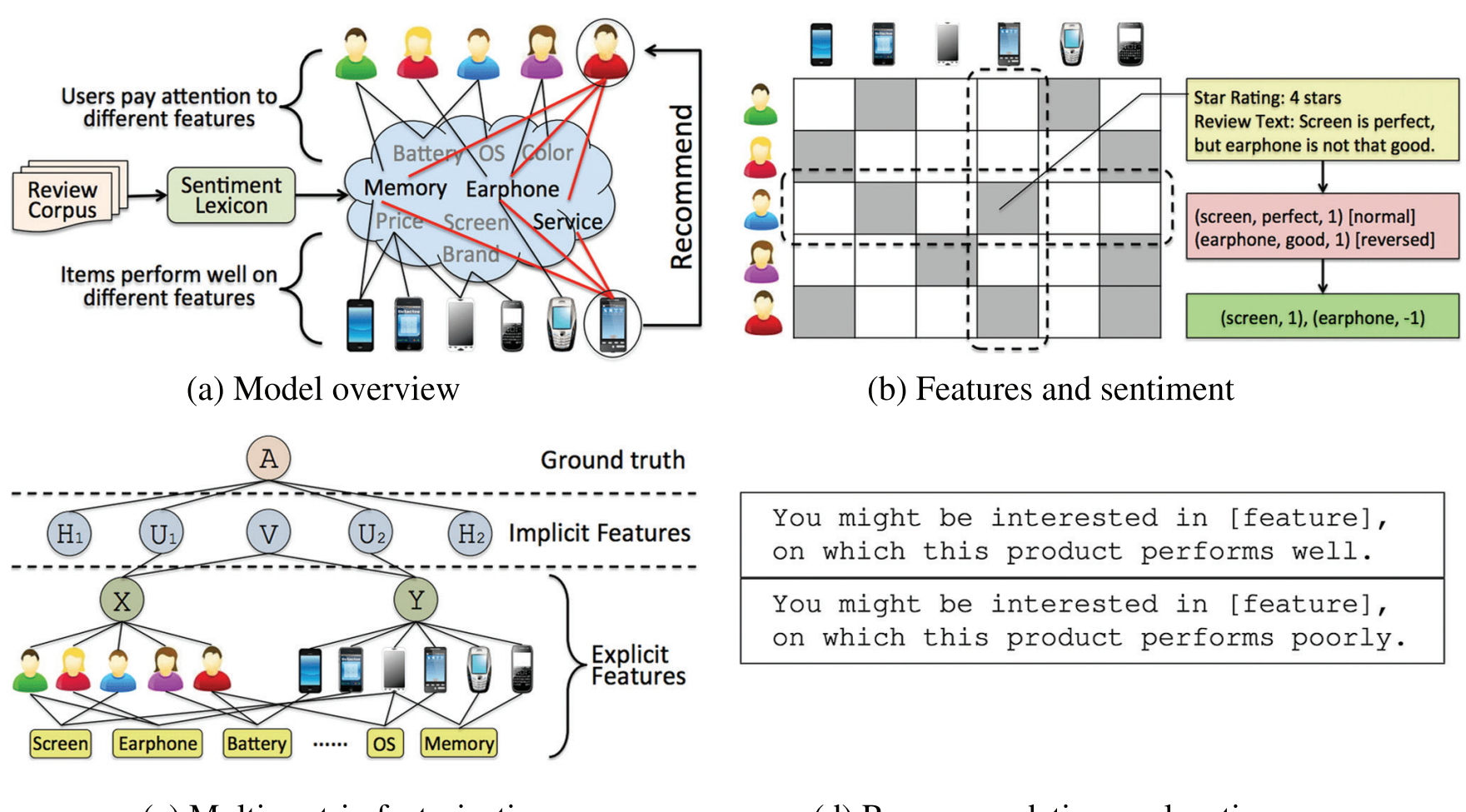

(a) Model overview

(b) Features and sentiment

(c) Multi-matrix factorization

(d) Recommendation explanations

**Figure 3.1:** Overview of the Explicit Factor Model. (a) The basic idea is to recommend products that perform well on the user's favorite features. (b) Each review (shaded block) is transformed into a set of product features accompanied by the user sentiment on the feature. (c) User attentions on different features constitute the user-feature attention matrix $X$, item qualities on features constitute the item-quality matrix $Y$, and these two matrices are collaborated to predict the rating matrix $A$. (d) The explicit product features can be used to generate personalized explanations.

Chen *et al.* (2016) further extended the EFM model to tensor factorization. In particular, the authors extracted product features from textual reviews and constructed the user-item-feature cube. Based on this cube, the authors conducted pair-wise learning to rank to predict user preferences on features and items, which helps to provide personalized recommendations. The model was further extended to consider multiple product categories simultaneously, which helps to alleviate the data sparsity problem, as shown in Figure 3.2(a).

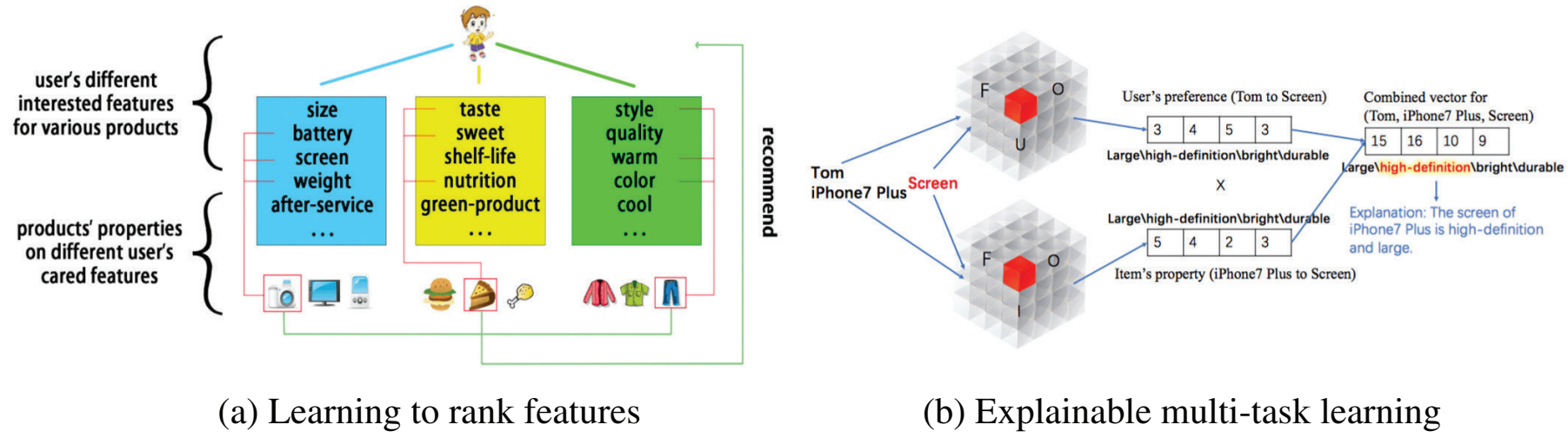

(a) Learning to rank features (b) Explainable multi-task learning

**Figure 3.2:** (a) Learning to rank features for explainable recommendation over multiple categories based on tensor factorization (Chen *et al.*, 2016). (b) User-feature-opinion tensor and item-feature-opinion tensor in explainable multi-task learning. Feature-level text-based explanation is generated by integrating these two tensors for a given user-item-feature tuple (Wang *et al.*, 2018b).

Wang *et al.* (2018b) further generalized MF-based explainable recommendation by multi-task learning over tensors. In particular, two companion learning tasks "user preference modeling for recommendation" and "opinionated content modeling for explanation" are integrated via a joint tensor factorization framework, as shown in Figure 3.2(b). The algorithm predicts not only a user's preference over a list of items (i.e., recommendations), but also how the user would appreciate a particular item at the feature level (i.e., opinionated textual explanation).

User preference distribution over the features could be different on different items, while the above methods assume each user has a global feature preference distribution. As an improvement, Chen *et al.* (2018b) propose an Attention-driven Factor Model (AFM), which learns and tunes the user attention distribution over features on different items. Meanwhile, users' attention distributions can also serve as explanations for recommendations.

To analyze the relationship between inputs (user history) and outputs (recommendations) in latent factor models, Cheng *et al.* (2019a)

adopted influence analysis in LFMs towards explainable recommendation. In particular, the authors incorporate interpretability into LFMs by tracing each prediction back to the model's training data, and further provide intuitive neighbor-style explanations for the predictions. We will provide more details of this work in the following section specifically devoted to model-agnostic and post-hoc explainable recommendation.

The features extracted from reviews can also be recommended to users as explanations. Bauman *et al.* (2017) proposed the Sentiment Utility Logistic Model (SULM), which extracts features (i.e., aspects) and user sentiments on these features. The features and sentiments are integrated into a matrix factorization model to regress the unknown ratings, which are finally used to generate recommendations. The proposed method provides not only item recommendations, but also feature recommendations for each items as explanations. For example, the method can recommend restaurants together with the most important aspects that the users can select, such as the time to go to a restaurant (e.g., lunch or dinner), and dishes to order (e.g., seafood). Qiu *et al.* (2016) and Hou *et al.* (2018) also investigated aspect-based latent factor models for recommendation by integrating ratings and reviews.

Latent factor models can also be integrated with other structured data for better recommendation and explainability, such as tree structures or graph structures. Tao *et al.* (2019a) proposed to tame latent factor models for explainability with factorization trees. The authors integrate regression trees to guide the learning of latent factor models for recommendation, and use the learned tree structure to explain the resulting latent factors. In particular, the authors build regression trees on users and items with user-generated reviews, and associate a latent profile to each node on the trees to represent users and items. With the growth of the regression tree, the latent factors are refined under the regularization imposed by the tree structure. As a result, we can track the creation of latent profiles by looking into the path of each factor on regression trees, which thus serves as an explanation for the resulting recommendations.

Researchers also investigated model-based approaches to generate relevant-user or relevant-item explanations, which provide explainable recommendations solely based on the user-item rating matrix (see Figure 3.3). Specifically, Abdollahi and Nasraoui (2016, 2017) described Explainable Matrix Factorization (EMF) for explainable recommendation. This model generates relevant-user explanations, where a recommended item is explained as *many users similar to you purchased this item*. To achieve this goal, the authors added an explainability regularizer into the objective function of matrix factorization. The explainability regularizer forces the user latent vector and item latent vector to be close to each other if a lot of the user's neighbors also purchased the item. In this way, the model naturally selects those commonly purchased items from a user's neighbors as recommendations, and meanwhile maintains high rating prediction accuracy. Liu *et al.* (2019) extended the idea by considering the fact that ratings in the user-item interaction matrix are missing *not* at random. The authors proposed an explainable probabilistic factorization model, which employs an influence mechanism to evaluate the importance of the users' historical data, so that the most related users and items can be selected to explain each predicted rating. Based on the learned influence scores, five representative user/item groups can be identified from data, including *influential users, impressionable users, independent users, popular items*, and *long-tail items*, which can be used to generate explanations such as *we recommend $x$ because an influential user $u$ liked the popular item $x$.*

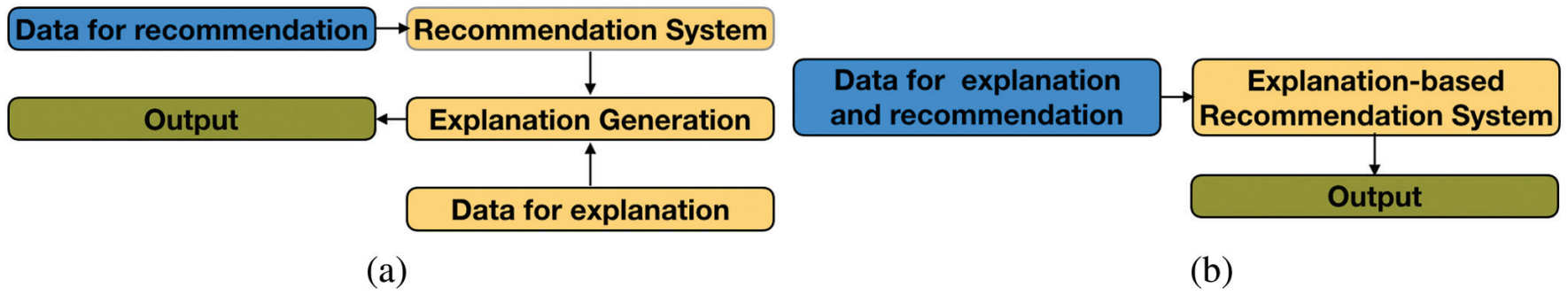

**Figure 3.3:** (a) Explainable recommendation with external data. (b) Explainable recommendation without external data support (Abdollahi and Nasraoui, 2016).

## 3.3 Topic Modeling for Explainable Recommendation

Based on available text information – especially the widely available textual reviews in e-commerce – topic modeling approach has also been widely adopted for explanations in recommender systems. In these approaches, users can usually be provided with intuitive explanations in the form of topical word clouds (McAuley and Leskovec, 2013; Wu and Ester, 2015; Zhao *et al.*, 2015). In this section, we review the related work that can be categorized into this approach.

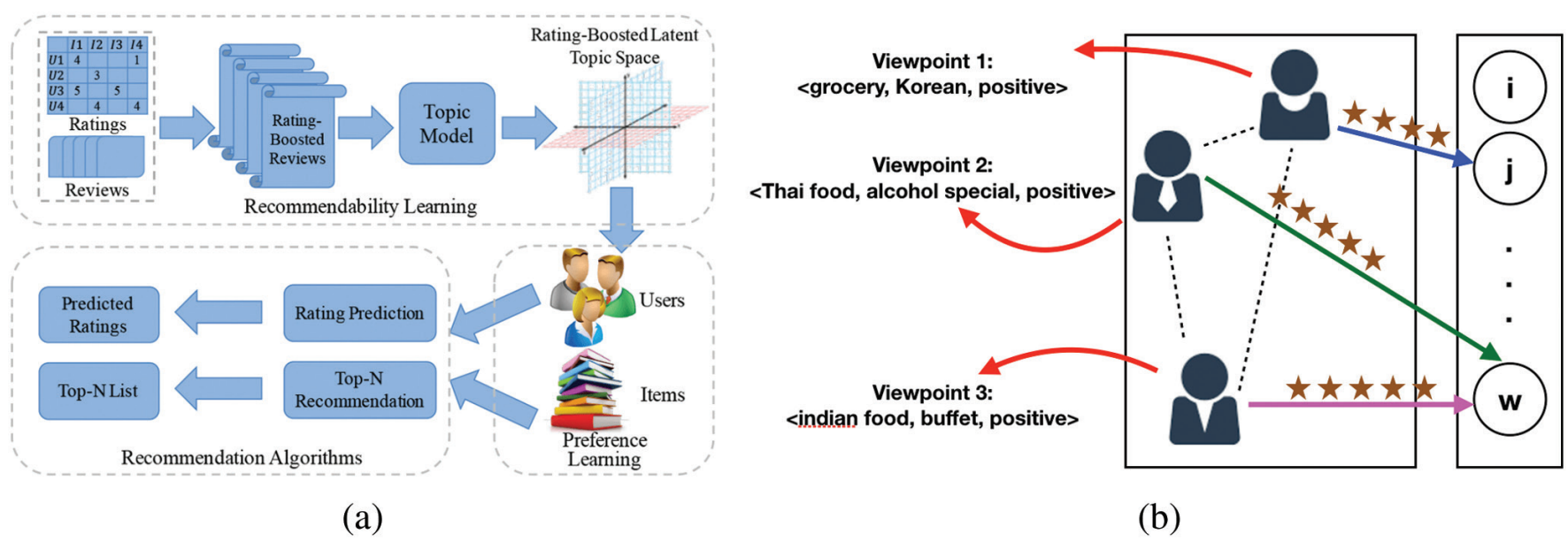

**Figure 3.4:** (a) Framework for understanding users and items based on ratings and reviews in the Rating-Boosted Latent Topics model. Users and items are embedded as recommendability vectors in the same space. They are further used by latent factor models for rating prediction and top-n recommendation (Tan *et al.*, 2016). (b) An example of trusted social relations, user reviews and ratings in a recommender system. Black arrows connect users with trusted social relations. "ThumpUp" symbols reflect the ratings of items. Concepts and topics have been highlighted in red and blue, respectively. Three viewpoints are represented in three different colors. A viewpoint is a mixture over a concept, a topic, and a sentiment (Ren *et al.*, 2017).

McAuley and Leskovec (2013) proposed to understand the hidden factors in latent factor models based on the hidden topics learned from reviews. The authors proposed the Hidden Factor and Topic (HFT) model, which bridges latent factor models and Latent Dirichlet Allocation (LDA). It links each dimension of the latent vector with a dimension of the LDA topic distribution vector. By considering reviews, the method improves rating prediction accuracy. Besides, by projecting each user's latent vector into the latent topic space in LDA, it helps to understand why a user made a particular rating on a target item. For example, we can detect the most significant topics that a user likes.

Following this idea, Tan *et al.* (2016) proposed to model item recommendability and user preference in a unified semantic space, as shown in Figure 3.4(a). In this model, each item is embedded as a topical recommendability distribution vector. Similarly, each user is embedded in the same topical recommendability space based on his/her historical ratings. The recommendability and preference distributions are, at last, integrated into a latent factorization framework to fit the ground truth. In this model, recommendation explanations can be derived by showing the topic words that have the most significant recommendability score. Cheng *et al.* (2019b) attempted to enrich topic models based on multi-modality information. In particular, the authors proposed a multi-modal aspect-aware topic modeling approach based on textual reviews and item images. The model learns user preferences and item features from different aspects, and also estimates the aspect importance of a user toward an item to provide explainable recommendations.

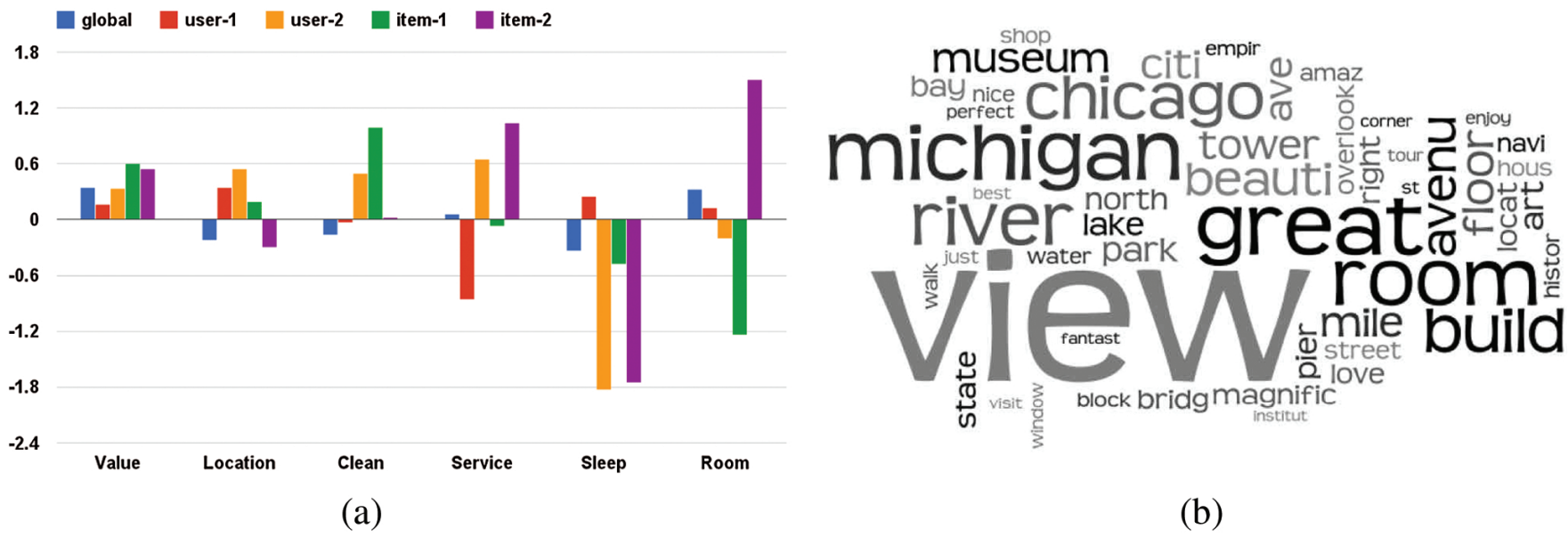

**Figure 3.5:** (a) The FLAME model learns each user's sentiments on different item aspects. (b) Show explanations as word clouds where the aspect size is proportional to its sentiment (Wu and Ester, 2015).

More generally, researchers also investigated other probabilistic graphic models beyond LDA for explainable recommendations. Wu and Ester (2015) studied the personalized sentiment estimation problem on item aspects. In particular, the authors proposed the FLAME model (Factorized Latent Aspect ModEl), which combines the advantages of collaborative filtering and aspect-based opinion mining. It learns each user's personalized preferences on different item aspects based on her reviews, as shown in Figure 3.5(a). Based on this, it predicts the user's aspect ratings on new items through collective intelligence. The proposed method showed improved performance on hotel recommendations on TripAdvisor. Further, for each hotel recommendation, it can provide a word cloud of the hotel aspects as an explanation, as shown in Figure 3.5(b), where the size of each aspect is proportional to the sentiment of the aspect.

Zhao *et al.* (2015) designed a probabilistic graphical model to integrate sentiment, aspect, and region information in a unified framework for improved recommendation performance and explainability in point-of-interest (POI) recommendation. The explanations are determined by each user's topical-aspect preference, which is similar to the topical clusters in McAuley and Leskovec (2013), but the difference is that the model can provide sentiment of each cluster for explanation purposes.

Ren *et al.* (2017) leveraged topic modeling for social explainable recommendations. Specifically, the authors proposed social-collaborative viewpoint regression (sCVR). A "viewpoint" is defined as a tuple of concept, topic, and sentiment label from both user reviews and trusted social relations, as shown in Figure 3.4(b), and the viewpoints are used as explanations. The authors proposed a probabilistic graphical model based on the viewpoints to improve prediction accuracy. Similar to the previous work, explanations are generated based on the user's favorite topics embedded in the viewpoints.

## 3.4 Graph-based Models for Explainable Recommendation

Many user-user or user-item relationships can be represented as graphs, especially in social network scenarios. In this section, we introduce how explainable recommendations can be generated based on graph learning approaches such as graph-based propagation and clustering. He *et al.* (2015) introduced a tripartite graph structure to model the user-item-aspect ternary relations for top-N recommendation, as shown in Figure 3.6, where an aspect is an item feature extracted from user reviews. The authors proposed TriRank, a generic algorithm for ranking the vertices in the tripartite graph, which applies smoothness and fitting constraints for node ranking and personalized recommendation. In this survey, explanations are attributed to the top-ranked aspects that match the target user and the recommended item.

Without using external information such as aspects, Heckel *et al.* (2017) conducted over-lapping co-clustering based on user-item bipartite graph for explainable recommendation. In each co-cluster, the users have similar interests, and the items have similar properties, as shown in Figure 3.7. Explanations are generated based on users' collaborative information, for example, in the form of "Item A is recommended to Client X with confidence $\alpha$, because Client X has purchased Item B, C, and D, while clients with similar purchase history (such as Clients Y and Z) also bought Item A". If a user-item pair falls into multiple co-clusters, we can thus generate multiple user-based and item-based explanations from each of the co-cluster.

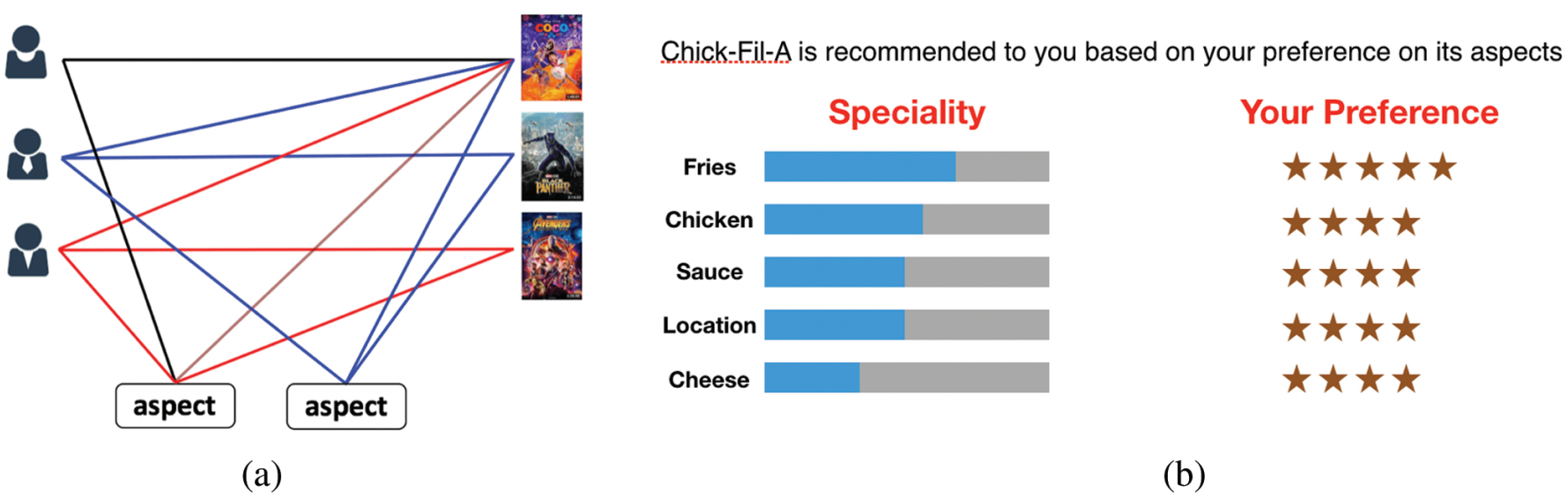

**Figure 3.6:** (a) An example tripartite structure of the TriRank algorithm. (b) A mock interface for showing the explanations of recommendation *Chick-Fil-A* to a user (He *et al.*, 2015).

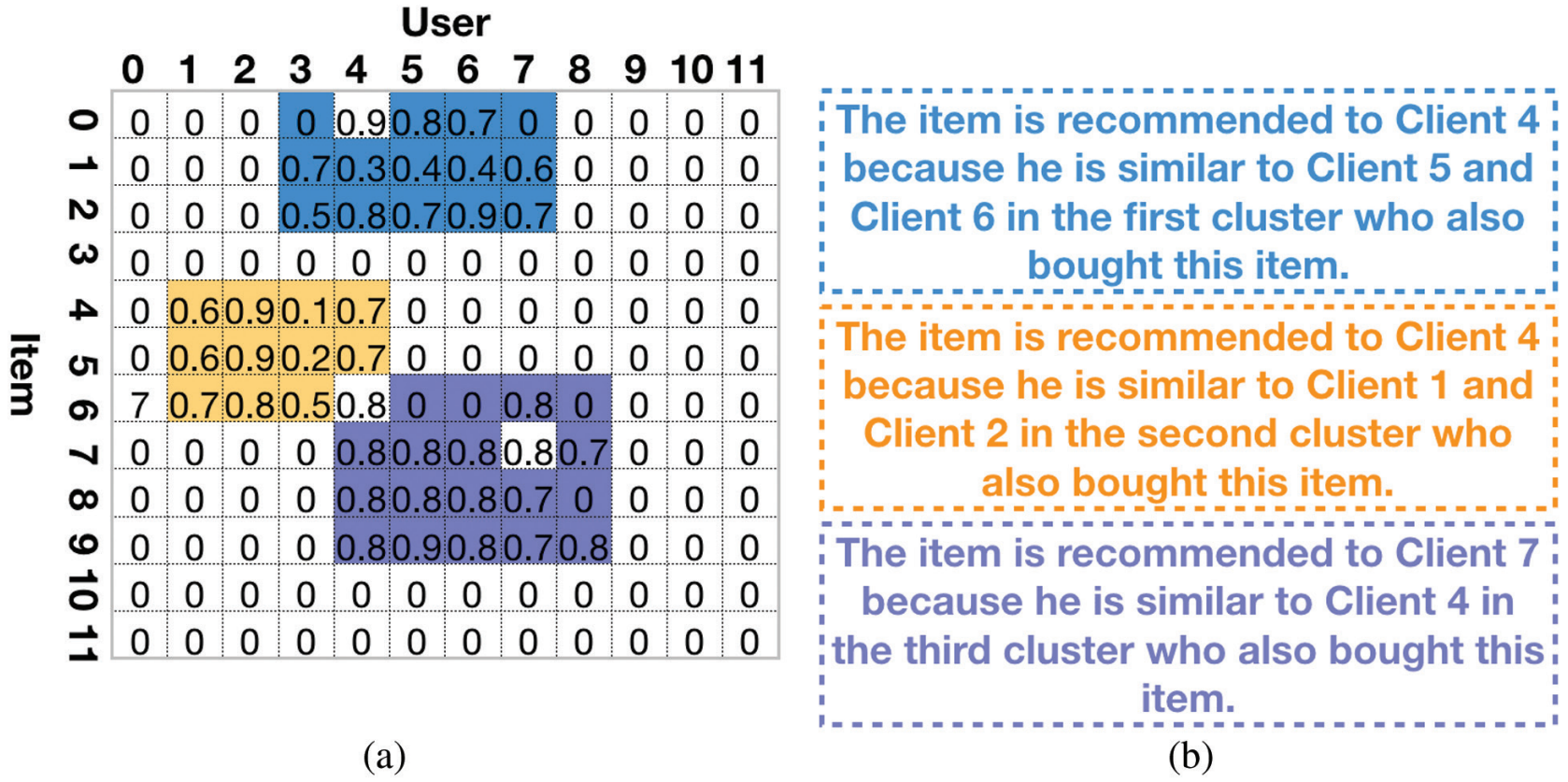

| Item \ User | 0 | 1 | 2 | 3 | 4 | 5 | 6 | 7 | 8 | 9 | 10 | 11 |
|---|---|---|---|---|---|---|---|---|---|---|---|---|
| 0 | 0 | 0 | 0 | 0 | 0.9 | 0.8 | 0.7 | 0 | 0 | 0 | 0 | 0 |
| 1 | 0 | 0 | 0 | 0.7 | 0.3 | 0.4 | 0.4 | 0.6 | 0 | 0 | 0 | 0 |
| 2 | 0 | 0 | 0 | 0.5 | 0.8 | 0.7 | 0.9 | 0.7 | 0 | 0 | 0 | 0 |
| 3 | 0 | 0 | 0 | 0 | 0 | 0 | 0 | 0 | 0 | 0 | 0 | 0 |
| 4 | 0 | 0.6 | 0.9 | 0.1 | 0.7 | 0 | 0 | 0 | 0 | 0 | 0 | 0 |
| 5 | 0 | 0.6 | 0.9 | 0.2 | 0.7 | 0 | 0 | 0 | 0 | 0 | 0 | 0 |
| 6 | 7 | 0.7 | 0.8 | 0.5 | 0.8 | 0 | 0 | 0.8 | 0 | 0 | 0 | 0 |
| 7 | 0 | 0 | 0 | 0 | 0.8 | 0.8 | 0.8 | 0.8 | 0.7 | 0 | 0 | 0 |
| 8 | 0 | 0 | 0 | 0 | 0.8 | 0.8 | 0.8 | 0.7 | 0 | 0 | 0 | 0 |
| 9 | 0 | 0 | 0 | 0 | 0.8 | 0.9 | 0.8 | 0.7 | 0.8 | 0 | 0 | 0 |
| 10 | 0 | 0 | 0 | 0 | 0 | 0 | 0 | 0 | 0 | 0 | 0 | 0 |
| 11 | 0 | 0 | 0 | 0 | 0 | 0 | 0 | 0 | 0 | 0 | 0 | 0 |

**Figure 3.7:** (a) Example of the overlapping user-item co-clusters identified by the OCuLaR algorithm (Heckel *et al.*, 2017). Colored elements are positive examples, and the white elements within the clusters are recommendations. (b) Based on the clustering results, the algorithm provides user-based and item-based explanations. For example, customers with history also purchased the recommended item.

Wang *et al.* (2018c) proposed a tree-enhanced embedding model for explainable recommendation to combine the generalization ability of embedding-based models and the explainability of tree-based models. In this model, the authors first employed a tree-based model to learn explicit decision rules. The decision rules are based on cross features extracted from side information. Then, the authors designed an embedding model that incorporates explicit cross features, and generalize to unseen user or item cross features based on collaborative learning. To generate explanations, an attention network is used to detect the most significant decision rules during the recommendation process.

Graph-based explainable recommendation is also frequently used in social recommendation scenarios, because social network is naturally a graph structure. For example, the UniWalk algorithm (Park *et al.*, 2018) introduced in Section 2 is a graph-based explainable recommendation algorithm. It exploits both ratings and the social network to generate explainable product recommendations. In this algorithm, a recommendation can be explained based on the target user's friends who have similar preferences on the graph.

## 3.5 Deep Learning for Explainable Recommendation

Recently, researchers have leveraged deep learning and representation learning for explainable recommendations. The deep explainable recommendation models cover a wide range of deep learning techniques, including CNN (Seo *et al.*, 2017; Tang and Wang, 2018), RNN/LSTM (Donkers *et al.*, 2017), attention mechanism (Chen *et al.*, 2018a), memory networks (Chen *et al.*, 2018c; Tao *et al.*, 2019b), capsule networks (Li *et al.*, 2019), and many others. They are also applied to different explainable recommendation tasks, such as rating prediction, top-n recommendation, and sequential recommendation. Based on natural language generation models, the system can even automatically generate explanation sentences instead of using explanation templates (Chen *et al.*, 2019a; Li *et al.*, 2017; Seo *et al.*, 2017). In this section, we will review deep learning approaches to explainable recommendation.

Seo *et al.* (2017) proposed to model user preferences and item properties using convolutional neural networks (CNNs) upon review text. It leans dual local and global attentions for explanation purposes, as shown in Figure 3.8. When predicting the user-item rating, the model selectively chooses review words with different attention weights. Based on the learned attention weights, the model can show which part of the review is more important for the output. Besides, the model can highlight the important words in the review as explanations to help users understand the recommendations.

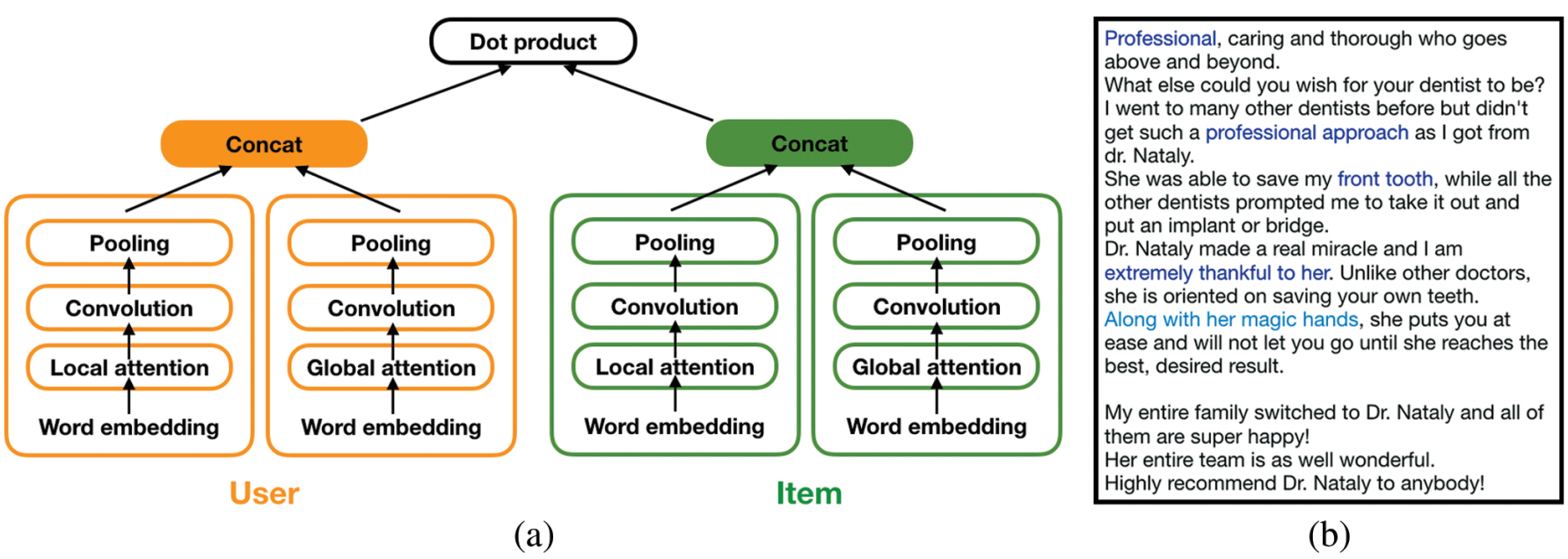

**Figure 3.8:** (a) The dual-attention architecture to extract user and item representations. A user document and an item document are fed into the user network (left) and item network (right). (b) The model generates attention scores for each review and highlights the high attention words as explanations (Seo *et al.*, 2017).

Similarly, Wu *et al.* (2019) combined the user-item interaction and review information in a unified framework. The user reviews are attentively summarized as content features, which are further integrated with the user and item embeddings to predict the final ratings. Lu *et al.* (2018a) presented a deep learning recommendation model that co-learns user and item information from ratings and reviews. It jointly optimizes a matrix factorization component (over ratings) and an attention-based GRU network (over reviews), so that features learned from ratings and reviews are aligned with each other. Lu *et al.* (2018b) further added a review discriminator based on adversarial sequence to sequence learning into the joint optimization framework, so that the generator can generate reviews for a user-recommendation pair as natural language explanation. In both (Wu *et al.*, 2019) and (Lu *et al.*, 2018a), the attention weights over review words are leveraged to explain the recommendations.

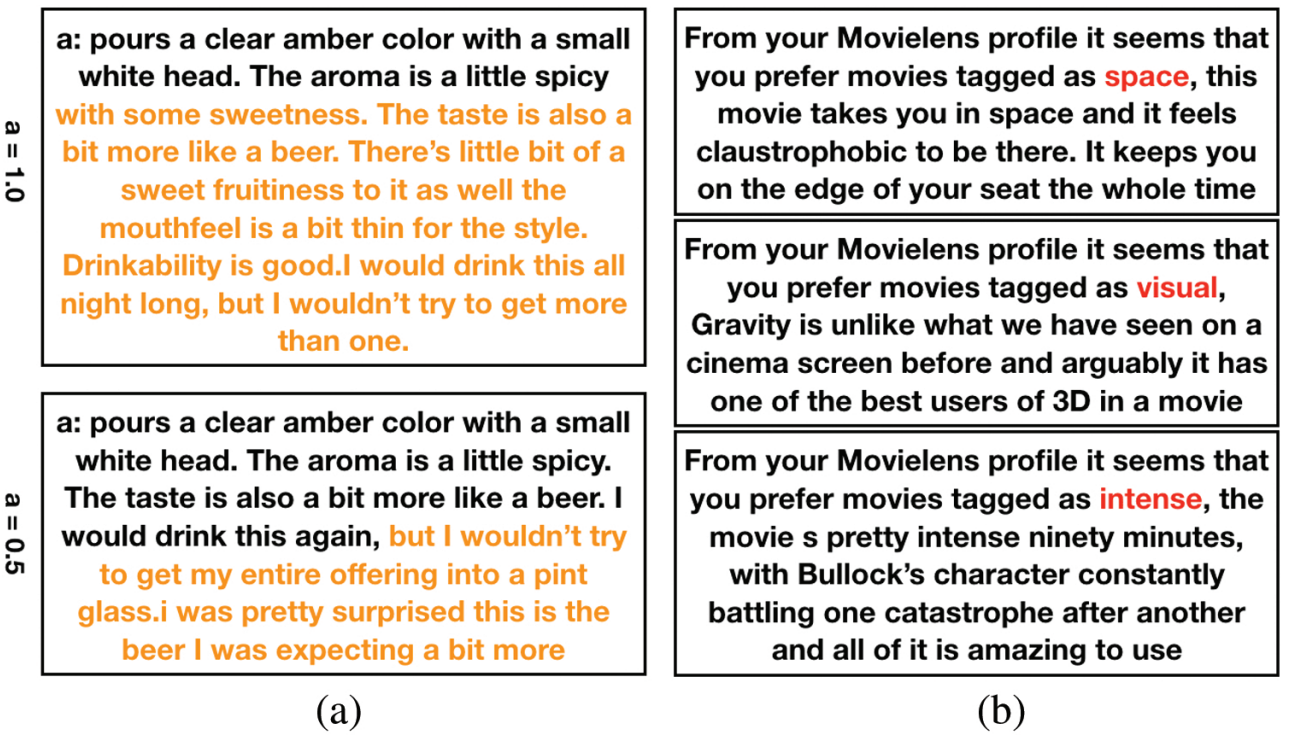

**Figure 3.9:** (a) Automatically generated textual reviews (explanations) based on natural language generation. Setting different model parameters will generate different explanations (Costa *et al.*, 2018). (b) Example natural language explanations for the movie "Gravity". Depending on the model of a user's interest, the system selects one crowd-sourced explanation for the user (Chang *et al.*, 2016).

Gao *et al.* (2019) developed a Deep Explicit Attentive Multi-view Learning Model (DEAML) for explainable recommendation, which aims to mitigate the trade-off between accuracy and explainability by developing explainable deep models. The basic idea is to build an initial network based on an explainable deep hierarchy (e.g., the Microsoft Concept Graph), and improve the model accuracy by optimizing key variables in the hierarchy (e.g., node importance and relevance). The model outputs feature-level explanations similar to Zhang *et al.* (2014a), but the features are attentively retrieved from an explicit feature hierarchy. The model is capable of modeling multi-level explicit features from noisy and sparse data, and shows highly usable explanations in industry-level applications. Instead of using existing features, Ma *et al.* (2019a) proposed to automatically learn disentangled features from data for recommendation, which not only improved the explainability but also enable users to better control the recommendation results.

Different from highlighting the review words as explanations, Costa *et al.* (2018) proposed a method for automatically generating natural language explanations based on character-level RNN. The model concatenates the user ratings into the input component as auxiliary information, so that the model can generate reviews according to the expected rating (sentiment). Different from many explainable recommendation models, whose explanation is generated based on predefined templates, the model can automatically generate explanations in a natural language manner. By choosing different parameters, the model can generate different explanations to attract users, as shown in Figure 3.9(a). Li *et al.* (2017) proposed a more comprehensive model to generate tips in review systems, where each tip is a short summarization sentence for a long review. The tips also serve as recommendation explanations. Chen *et al.* (2019a) integrated the natural language generation approach and the feature word approach, and proposed a topic sensitive generation model to generate explanations about particular feature words. To some extent, the model can control the item aspect that the generated explanation talks about. Inspired by human's information-processing model in cognitive psychology, Chen *et al.* (2019d) developed an encoder-selector-decoder architecture for explainable recommendation, which exploits the correlations between the recommendation task and the explanation task through co-attentive multi-task learning. The model enhances the accuracy of the recommendation task, and generates linguistic explanations that are fluent, useful, and highly personalized.

Chang *et al.* (2016) proposed another approach to generating natural language explanations, which is based on human users and crowdsourcing. The authors argued that algorithm generated explanations can be overly simplistic and unconvincing, while humans can overcome these limitations. Inspired by how people explain word-of-mouth recommendations, the authors designed a process to combine crowdsourcing and computation for generating explanations. They first extract the topical aspects of movies based on an unsupervised learning approach, and then, generate natural language explanations for the topical aspects. More specifically, the authors collect relevant review quotes for each aspect, and then ask crowd workers to synthesize the quotes into explanations. Finally, the authors model users' preferences based on their activities and present explanations in a personalized fashion (Figure 3.9(b)). Controlled experiments with 220 MovieLens users were conducted to evaluate the efficiency, effectiveness, trust, and satisfaction of the personalized natural language explanations, in comparison with personalized tag-based explanations.

Instead of generating explanations, Chen *et al.* (2018a) selects appropriate user reviews as explanations. The authors designed an attention mechanism over the user and item reviews for rating prediction. In this research, the authors believe that reviews written by others are critical reference information for a user to make decisions in e-commerce. However, the huge amount of reviews for each product makes it difficult for consumers to examine all the reviews to evaluate a product. As a result, selecting and providing high-quality reviews for each product is an important approach to generate explanations. Specifically, the authors introduced an attention mechanism to learn the usefulness of reviews. Therefore, highly-useful reviews can be adopted as explanations, which help users to make faster and better decisions.

Chen *et al.* (2019b) proposed visually explainable recommendation by jointly modeling visual images and textual reviews. It highlights the image region-of-interest for a user as recommendation explanations, as shown in Figure 2.10. By jointly modeling images and reviews, the proposed model can also generate a natural language explanation to accompany the visual explanations by describing the highlighted regions. With the natural language explanations, users can better understand why the particular image regions are highlighted as visual explanations.

With the advantage of reasoning over explicit memory slots, memory networks have been explored in explainable recommendation tasks. For example, Chen *et al.* (2018c) studied explainable sequential recommendation based on memory networks. It considers each item in a user's interaction history as a memory slot, and develops an attention mechanism over the slots to predict the subsequent user behaviors. Explanations are provided by showing how and which of the user's previous item(s) influenced the current prediction. The authors further proposed Dynamic Explainable Recommendation (Chen *et al.*, 2019c) based on time-aware gated recurrent units. Tao *et al.* (2019b) proposed a Log2Intent framework for interpretable user modeling. It focuses on modeling user behaviors, as well as predicting and interpreting user intents from the unstructured software log-trace data. Technically, the authors incorporate auxiliary knowledge with memory networks for sequence to sequence modeling. The attention mechanism produces attended annotations over memory slots to interpret the user log data.

Li *et al.* (2019) developed a capsule network approach to explainable recommendation. It considers an "item aspect – user viewpoint" pair as a logic unit, which is used to reason the user rating behaviors. The model discovers the logic units from reviews and resolves their sentiments for explanations. A sentiment capsule architecture with a Routing by Bi-Agreement mechanism is proposed to identify the informative logic units for rating prediction, while the informativeness of each unit helps to produce explanations for the predictions. Developing capsule logic units for explainable reasoning shows a promising approach towards explainable recommendation systems.

We should note that the scope and literature of deep learning for explainable recommendation is not limited to the research introduced in this subsection. Except for deep learning over text or image for explainable recommendations, research efforts in the explainable/interpretable machine learning area are also reflected in explainable recommendation systems. Though they also belong to deep learning for explainable recommendations, many of them are better classified into other explainable recommendation approaches such as model-agnostic or post-hoc methods. We will introduce these methods in the following subsections.

Another important yet less explored question is the fidelity of the deep explainable models. Deep learning models are usually complex in nature, and sometimes it could be difficult to decide if the explanations provided by the model truly reflect the real mechanism that generated the recommendations or decisions. The community has different opinions on the explanation fidelity problem. For example, attention mechanism is a frequently used technique to design explainable decision making models. However, Jain and Wallace (2019) argued that standard attention modules do not provide meaningful explanations and should not be treated as though they do, while Wiegreffe and Pinter (2019) later challenged many of the assumptions underlying the prior work, arguing that such a claim depends on one's definition of explanation, and showed that the prior work does not disprove the usefulness of attention mechanisms for explainability. Overall, the explainability of deep models is still an important open problem to explore, and more advanced explanation models are needed to understand the behavior of neural networks. We will further discuss this problem in the following section of evaluating explainable recommendations.

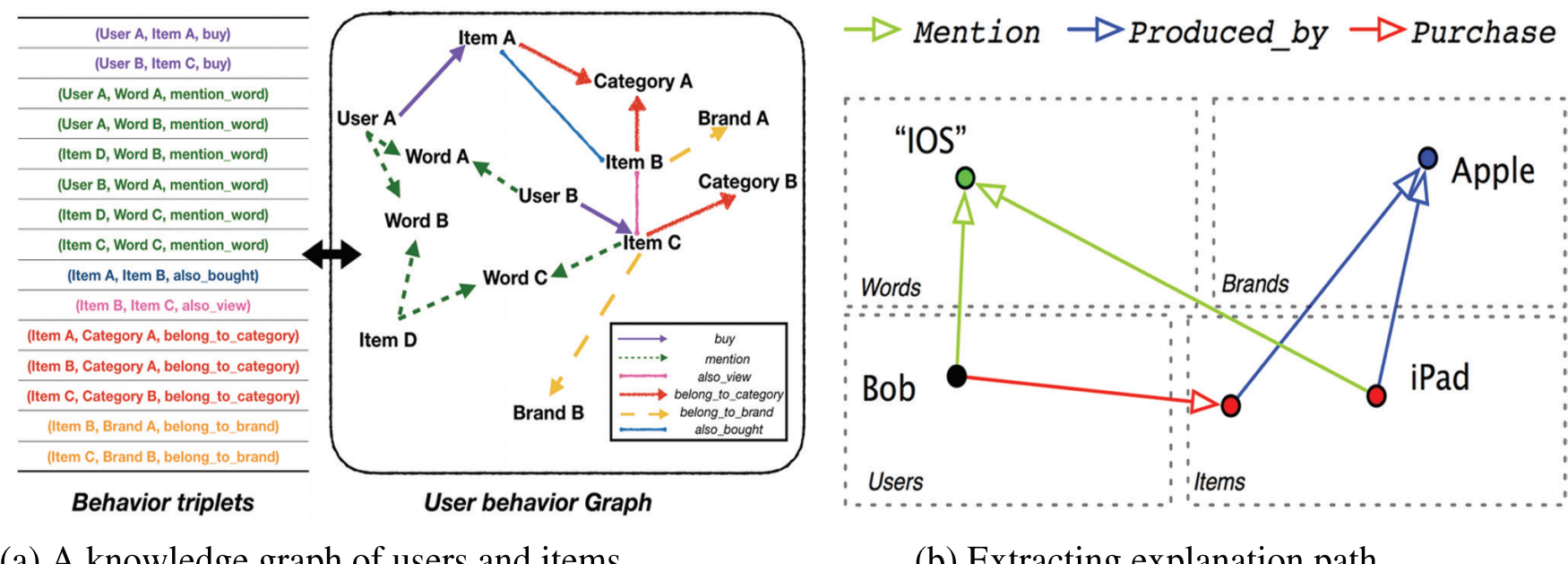

(a) A knowledge graph of users and items

(b) Extracting explanation path

**Figure 3.10:** (a) The user-item knowledge graph constructed for Amazon product domain. On the left is a set of triplets of user behaviors and item properties, and on the right is the corresponding graph structure. The knowledge graph contains different types of relations such as *purchase, mention, also bought, also view, category*, and *brand*. (b) Explanation paths between a user Bob and a recommended item iPad in the product knowledge graph. Bob and iPad can be connected through the commonly mentioned word *iOS* or the common brand *Apple* from Bob's already purchased products.

## 3.6 Knowledge Graph-based Explainable Recommendation

Knowledge graph (KG, or knowledge base) contains rich information about the users and items, which can help to generate intuitive and more tailored explanations for the recommended items. Recently, researchers have been exploring knowledge graphs for explainable recommendations.

Catherine *et al.* (2017) proposed a method to jointly rank items and knowledge graph entities using a Personalized PageRank procedure, which produces recommendations together with their explanations. The paper works on the movie recommendation scenario. It produces a ranked list of entities as explanations by jointly ranking them with the corresponding movies.

Different from Catherine *et al.* (2017) that adopts rules and programs on KG for explainable recommendations, Ai *et al.* (2018) proposed to adopt KG embeddings for explainable recommendation, as shown in Figure 3.10. The authors constructed a user-item knowledge graph, which contains various user, item, and entity relations, such as "user *purchase* item", and "item *belong to* category". KG embeddings are learned over the graph to obtain the embeddings of each user, item, entity, and relation. To decide recommendations for a user, the model finds the most similar item under the *purchase* relation. Besides, explanations can be provided by finding the shortest path from the user to the recommended item through the KG. By incorporating explicit user queries, the model can be further extended to conduct explainable search over knowledge graphs (Ai *et al.*, 2019).

Wang *et al.* (2018a) proposed the Ripple Network, an end-to-end framework to incorporate KG into recommender systems. Similar to ripples propagating on the surface of the water, the Ripple Network stimulates the propagation of user preferences over the knowledge entities. It automatically and iteratively extends a user's potential interests through the links in the KG. The multiple "ripples" activated by a user's historically clicked items are thus superposed to form the preference distribution of the user for a candidate item, which can be used to predict the final click probability. Explanations can also be provided by finding a path from the user and the recommended item over the knowledge graph.

Huang *et al.* (2018) leveraged KG for recommendation with better explainability in a sequential recommendation setting. The authors bridged Recurrent Neural Network (RNN) with Key-Value Memory Networks (KV-MN) for sequential recommendation. The RNN component is used to capture a user's sequential preference on items, while the memory network component is enhanced with knowledge of items to capture the users' attribute-based preferences. Finally, the sequential preferences, together with the attribute-level preferences, are combined as the final representation of user preference. To explain the recommendations, the model detects those attributes that are taking effect when predicting the recommended item. For a particular music recommendation as an example, it can detect whether the *album* attribute is more important or the *singer* attribute is more important, where the attributes come from an external knowledge graph. The model further enhances the explainability by providing value-level interpretability, i.e., suppose we already know that some attributes (e.g., album) play a critical role in determining the recommendation, the model further predicts the importance of different values for that attribute. Huang *et al.* (2019) further incorporated multi-modality knowledge graph for explainable sequential recommendation. Different from conventional item-level sequential modeling methods, the proposed method captured user dynamic preferences on user-item interaction-level by modeling the sequential interactions over knowledge graphs.

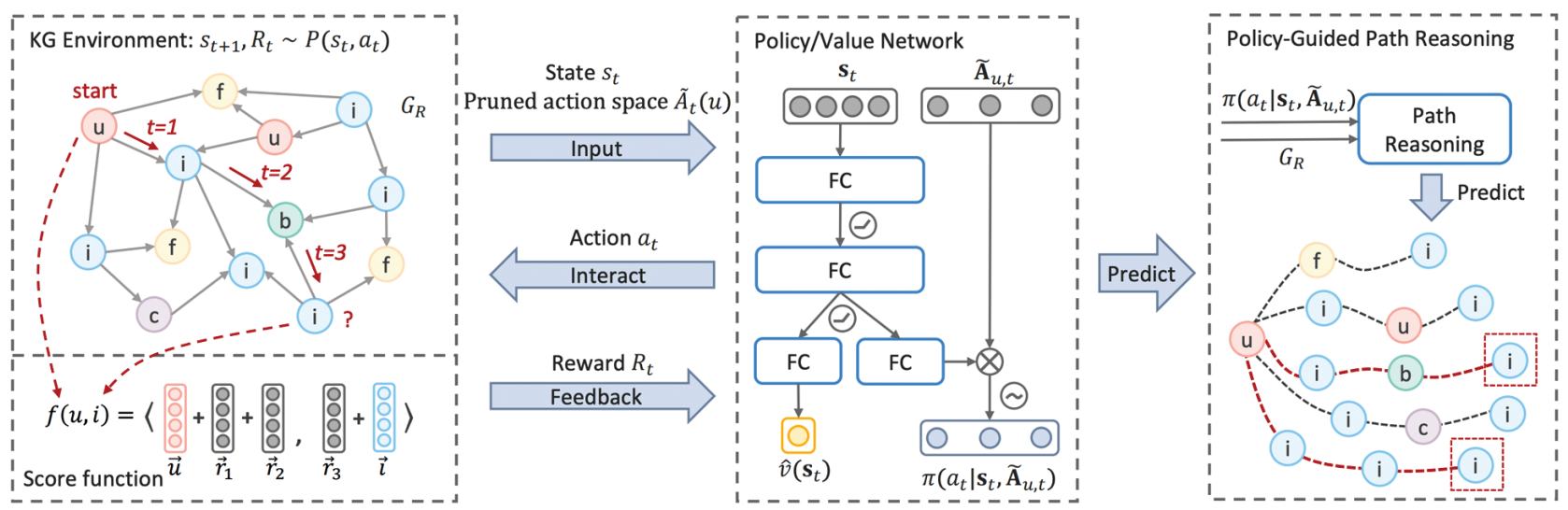

**Figure 3.11:** Policy-guided path reasoning for explainable recommendation. The algorithm aims to learn a policy that navigates from a user to potential items of interest by interacting with the knowledge graph environment. The trained policy is then adopted for the path reasoning phase to make recommendations to the user, while the path serves the explanation for a recommendation (Xian *et al.*, 2019).

To combine the power of machine learning and inductive rule learning, Ma *et al.* (2019b) proposed a joint learning framework to integrate explainable rule induction in KG with a rule-guided neural recommendation model. The framework encourages two modules to complement each other in generating explainable recommendations. One module is based on inductive rules mined from item knowledge graphs. The rules summarize common multi-hop relational patterns for inferring the item associations, and they provide human-readable explanations for model prediction. The second module is a recommendation module, which is augmented by the induced rules. The KG inductive rules are translated into explanations, which connect the recommended item with the user's purchase history.

Real-world knowledge graphs are usually huge. Enumerating all the paths between a user-item node pair for similarity calculation is usually computationally prohibitive. To solve the problem, Xian *et al.* (2019) proposed a reinforcement reasoning approach over knowledge graphs for explainable recommendations, as shown in Figure 3.11. The key idea is to train a reinforcement learning agent for pathfinding. In the training stage, the agent starts from a user and is trained to reach the correct items with high rewards. In the inference stage, the agent will thus directly walk to correct items for recommendations, without having to enumerate all the paths between user-item pairs.

Knowledge graphs can also help to explain a blank-box recommendation model. Zhang *et al.* (2020) proposed a knowledge distillation approach to explain black-box models for recommendation. The authors proposed an end-to-end joint learning framework to combine the advantages of embedding-based recommendation models and path-based recommendation models. Given an embedding-based model that produces black-box recommendations, the proposed approach interprets its recommendation results based on differentiable paths on knowledge graphs; the differentiable paths, on the other hand, regularize the black-box model with structured information encoded in knowledge graph for better performance.

## 3.7 Rule Mining for Explainable Recommendation

Rule mining approaches are essential for recommendation research. They usually have special advantages for explainable recommendations, because in many cases, they can generate very straightforward explanations for users. The most frequently used rule mining technique for explainable recommendation is association rule mining (Agrawal *et al.*, 1993, 1994). A very classic example is the "beer-diaper" recommendation originated from data mining research.

For example, Mobasher *et al.* (2001) leveraged association rule mining for efficient web page recommendation at large-scale. Cho *et al.* (2002) combined decision tree and association rule mining for a web-based shop recommender system. Smyth *et al.* (2005) adopted apriori association rule mining to help calculate item-item similarities, and applied association rule mining for the conversational recommendation task. Sandvig *et al.* (2007) studied the robustness of collaborative recommendation algorithms based on association rule mining. Zhang *et al.* (2015a) defined a sequence of user demands as a web browsing task, by analyzing user browsing logs, they leveraged frequent pattern mining for task-based recommendations. More comprehensively, Amatriain and Pujol (2015) provided a survey of data mining for personalized recommendation systems.

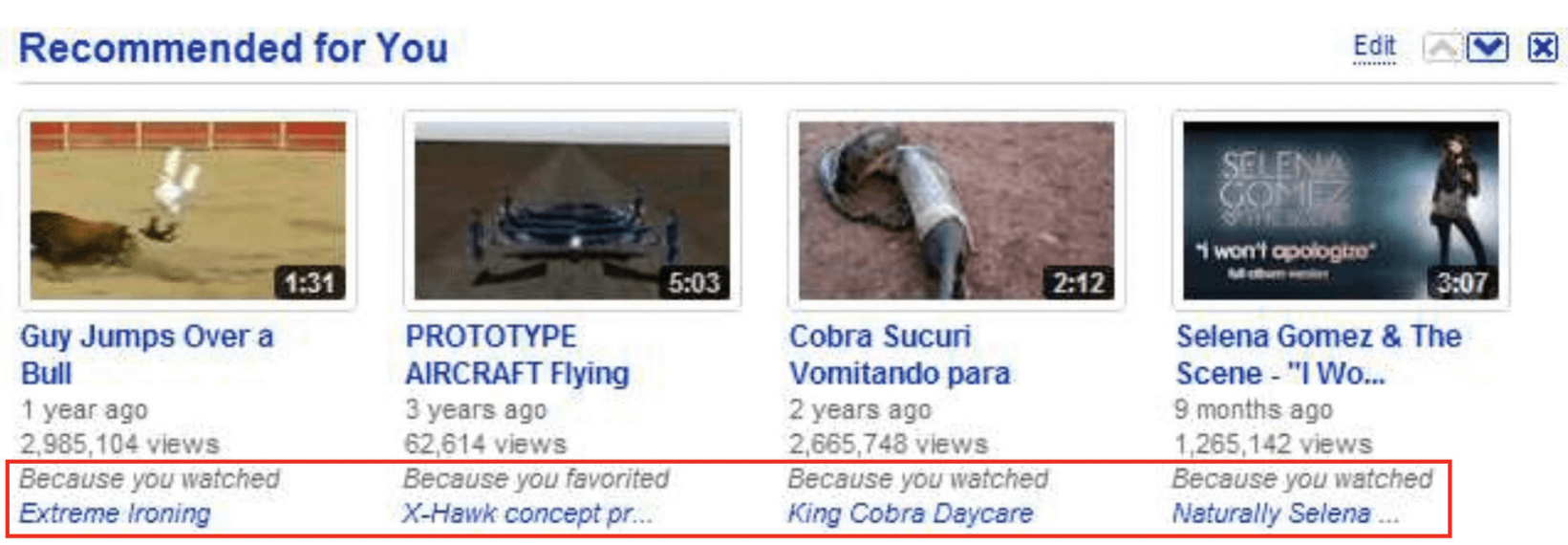

**Figure 3.12:** A screenshot of the "personalized recommendation" module on the YouTube home page. The boxed explanations are generated based on association rule mining (Davidson *et al.*, 2010).

In terms of explainable recommendation, Lin *et al.* (2000, 2002) investigated association rules for recommendation systems. In particular, the authors proposed a "personalized" association rule mining technique, which extracts association rules for a target user. The associations between users and items are employed to make recommendations, which are usually self-explainable by the association rules that produced them, for example, "90% of the articles liked by user A and user B are also liked by user C".

Davidson *et al.* (2010) introduced the YouTube Video Recommendation System. The authors considered the sessions of user watch behaviors on the site. For a given period (usually 24 hours), the authors adopted association rule mining to count how often each pair of videos $(v_i, v_j)$ were co-watched within the same session, which helps to calculate the relatedness score for each pair of videos. To provide personalized recommendations, the authors consider a video seed set for each user, which includes videos the user watched recently, and videos the user explicitly favorited, liked, rated, or added to playlists. Related videos of these seed videos are recommendations, while the seed video, as well as the association rules that triggered the recommendation, are taken as explanations, as shown in Figure 3.12.

Recently, Balog *et al.* (2019) proposed a set-based approach for transparent, scrutable, and explainable recommendations. Please note that although we discuss this work in this section of rule mining for explainable recommendation, the proposed approach is a framework that can be generalized to machine learning models depending on how item priors are estimated. The proposed model assumes that user preferences can be characterized by a set of tags or keywords. These tags may be provided by users (social tagging) or extracted automatically. Given explicit ratings of specific items, it infers set-based preferences by aggregating over items associated with a tag. This set-based user preference model enables us to generate item recommendations transparently and provide sentence-level textual explanations. A significant advantage of this explainable recommendation model is that it provides *scrutability* by letting users provide feedback on individual sentences. Any change to the user's preferences has an immediate impact, thereby endowing users with more direct control over the recommendations they receive.

## 3.8 Model Agnostic and Post Hoc Explainable Recommendation

Sometimes the recommendation mechanism may be too complex to explain. In such cases, we rely on post-hoc or model-agnostic approaches to explainable the recommendations. In these methods, recommendations and explanations are generated from different models – an explanation model (independent from the recommendation mechanism) provides explanations for the recommendation model after the recommendations have been provided (thus "post-hoc").

For example, in many e-commerce systems, the items are recommended based on very sophisticated hybrid models, but after an item is recommended, we can provide some simple statistical information as explanations, such as "70% of your friends bought this item". Usually, we pre-define several possible explanation templates based on data mining methods, such as frequent itemset mining and association rule mining, and then decide which explanation(s) to display based on post-hoc statistics such as maximum confidence. It should be noted that just because the explanations are post-hoc does not mean that they are fake, i.e., the statistical explanations should be true information, they are just decoupled from the recommendation model.

Peake and Wang (2018) provided an association rule mining approach to post-hoc explainable recommendations. The authors treated an arbitrary recommendation model – in this paper, a matrix factorization model – as a black box. Given any user, the recommendation model takes the user history as input and outputs the recommendations. The input and output constitute a transaction, and transactions of all users are used to extract association rules. The association rules can then be used to explain the recommendations produced by the black-box model – if an item recommended by the black-box model can also be recommended out of the association rules, we thus say that this item is explainable by the rules, such as "$\{X \Rightarrow Y\}$: Because you watched X, we recommend Y". The authors also adopted *fidelity* to evaluate the post-hoc explanation model, which shows what percentage of items can be explained by the explanation model.

Singh and Anand (2018) studied the post-hoc explanations of learning-to-rank algorithms in terms of web search. In this work, the authors focused on understanding the ranker decisions in a model agnostic manner, and the rankings explainability is based on an interpretable feature space. Technically, the authors first train a blackbox ranker, and then use the ranking labels produced by the ranker as secondary training data to train an explainable tree-based model. The tree-based model is the post-hoc explanation model to generate explanations for the ranking list. In this sense, Peake and Wang (2018) can be considered as a point-wise post-hoc explanation model, while Singh and Anand (2018) is a pair-wise post-hoc explanation model.

In general machine learning context, a prominent idea of model-agnostic explanation is using simple models to approximate a complex model around a sample, so that the simple models help to understand the complex model locally. For example, Ribeiro *et al.* (2016) proposed LIME (Local Interpretable Model-agnostic Explanation), which adopts sparse linear models to approximate a complex (non-linear) classifier around a sample, and the linear model can thus explain to us which feature(s) of the sample contributed to its predicted label. Singh and Anand (2019) extended the idea to explore the local explainability of ranking models. In particular, the authors converted the ranking problem over query-document pairs into a binary classification problem over relevant/irrelevant classes, and employed the LIME framework to explain the ranking model using linear SVM models. Coefficients of the linear model thus explain to us which words of a document are strong indicators of relevance.

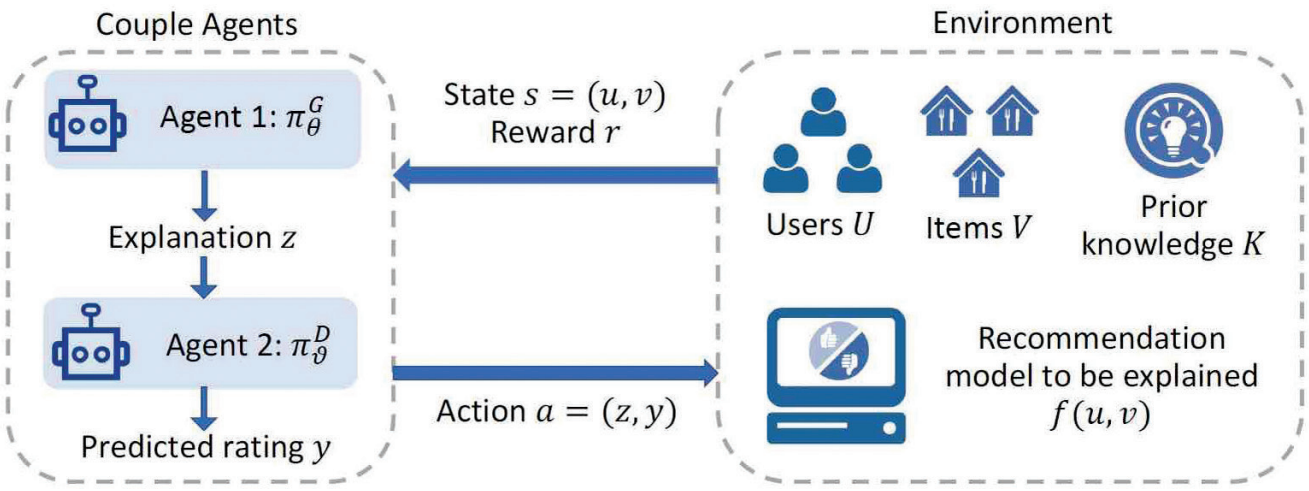

**Figure 3.13:** A reinforcement learning framework for generating recommendation explanations. The coupled agents learn to select the explanations that best approximate the predictions made by the recommendation model (Wang *et al.*, 2018d).

McInerney *et al.* (2018) developed a bandit approach to explainable recommendation. The authors proposed that users would respond to explanations differently and dynamically, and thus, a bandit-based approach for exploitation-exploration trade-off would help to find the best explanation orderings for each user. In particular, they proposed methods to jointly learn which explanations each user responds to, which are the best contents to recommend for each user, and how to balance exploration with exploitation to deal with uncertainty. Experiments show that explanations affect the way users respond to recommendations, and the proposed method outperforms the best static explanations ordering. This work shows that just as exploitation-exploration is beneficial to recommendation tasks, it is also beneficial to explanation tasks.

Wang *et al.* (2018d) proposed a model-agnostic reinforcement learning framework to generate sentence explanations for any recommendation model (Figure 3.13). In this design, the recommendation model to be explained is a part of the environment, while the agents are responsible for generating explanations and predicting the output ratings of the recommendation model based on the explanations. The agents treat the recommendation model as a black box (model-agnostic) and interact with the environment. The environment rewards the agents if they can correctly predict the output ratings of the recommendation model (model-explainability). Based on some prior knowledge about desirable explanations (e.g., the desirable length), the environment also rewards the agents if the explanations have good presentation quality (explanation quality control). The agents learn to generate explanations with good explainability and presentation quality by optimizing the expected reward of their actions. In this way, the recommendation model reinforces the explanation model towards better post-hoc explanations.

Cheng *et al.* (2019a) contributed mathematical understandings to post-hoc explainable recommendations based on influence analysis. Influence functions, which stem from robust statistics, have been used to understand the effect of training points on the predictions of black-box models. Inspired by this, the authors propose an explanation method named FIA (Fast Influence Analysis), which helps to understand the prediction of trained latent factor models by tracing back to the training data with influence functions. They presented how to employ influence functions to measure the impact of historical user-item interactions on the prediction results of LFMs, and provided intuitive neighbor-style explanations based on the most influential interactions.

Overall, post-hoc explainable recommendation approaches attempt to develop an explanation model to explain a black-box prediction model. Though the explanations may not strictly follow the exact mechanism that produced the recommendations (i.e., explanation fidelity may be limited), they have advantages in the flexibility to be applied in many different recommendation models.

## 3.9 Summary

In this section, we introduced different machine learning models for explainable recommendations. We first provided an overview of machine learning for recommender systems, and then, we focused on different types of explainable recommendation methods, including matrix/tensor-factorization approaches, topic modeling approaches, graph-based approaches, deep learning approaches, knowledge-based approaches, rule mining approaches, and post-hoc/model-agnostic approaches.

Explainable recommendation research can consider the explainability of either the recommendation methods or the recommendation results, corresponding to model-intrinsic and model-agnostic approaches (Lipton, 2018). It is interesting to see that both modeling philosophies have their roots in human cognitive science (Miller, 2019). Sometimes, human-beings make decisions based on careful logical reasoning, so they can clearly explain how a decision is made by showing the reasoning process step by step (Lipton, 2018). In this case, the decision model is transparent and the explanations can be naturally provided. Other times, people make intuition decisions first, and then "seek" an explanation for the decision, which belongs to the post-hoc explanation approach (Miller, 2019). It is difficult to say which research philosophy towards explainable recommendation – and explainable AI in a broader sense – is the correct approach (maybe both). Answering this question requires significant breakthroughs in human cognitive science and our understanding about how the human brain works.

# 4

# Evaluation of Explainable Recommendation

In this section, we provide a review of the evaluation methods for explainable recommendations. It would be desirable if an explainable recommendation model can achieve comparable or even better recommendation performance than conventional "non-explainable" methods, and meanwhile, achieve better explainability.

To evaluate the recommendation performance, we can adopt the same measures as evaluating conventional recommendation algorithms. For rating prediction tasks, we can use mean absolute error (MAE) or root mean square error (RMSE), while for top-n recommendation, we can adopt standard ranking measures such as precision, recall, F-measure, and normalized discounted cumulative gain (NDCG). We can also conduct online evaluations with real users based on online measures such as click-through rate (CTR) and conversion rate, which are frequently used to evaluate ranking performance.

In this section, we mainly focus on the evaluation of explanations. Similarly, explanations can also be evaluated both online and offline. Usually, offline evaluation is easier to implement, since online evaluation and user studies would depend on the availability of data and users in real-world systems, which are not always accessible to researchers.

As a result, online evaluation is encouraged but not always required for explainable recommendation research.

## 4.1 User Study

A straightforward approach to evaluating explanations is through user study based on volunteers or paid experiment subjects. The volunteers or paid subjects can either be directly recruited by the researchers or through online crowdsourcing platforms such as Amazon Mechanical Turk[1] and CrowdFlower.[2] Usually, the study will design some questions or tasks for the subjects to answer or complete, and conclusions will be derived from the responses of the subjects (Kittur *et al.*, 2008).

For example, Herlocker *et al.* (2000) studied the effectiveness of different explanation styles in user-based collaborative filtering. In this research, the study was performed as a survey, and study participants were presented with the following hypothetical situation:

> Imagine that you have $7 and a free evening coming up. You are considering going to the theater to see a movie, but only if there is a movie worth seeing. To determine if there is a movie worth seeing, you consult MovieLens for a personalized movie recommendation. MovieLens recommends one movie, and provides some justification.

Each participant was then provided with 21 individual movie recommendations, each with a different explanation component (see Figure 2.2 for two examples), and asked to rate on a scale of one to seven how likely they would be to go and see the movie. The subject's average responses on each explanation are thus calculated to evaluate the effectiveness the explanation.

Vig *et al.* (2009) conducted a user study for four explanation interfaces based on the MovieLens website, as shown in Figure 4.1. Subjects complete an online survey, in which they evaluate each interface about how well it helps them to: 1) understand why an item is recommended (justification), 2) decide if they like the recommended item (effectiveness), and 3) determine if the recommended item matches their mood (mood compatibility). The survey responses help the authors to conclude the role of tag preference and tag relevance in promoting justification, effectiveness, and mood compatibility.

[1] https://www.mturk.com
[2] https://www.crowdfower.com

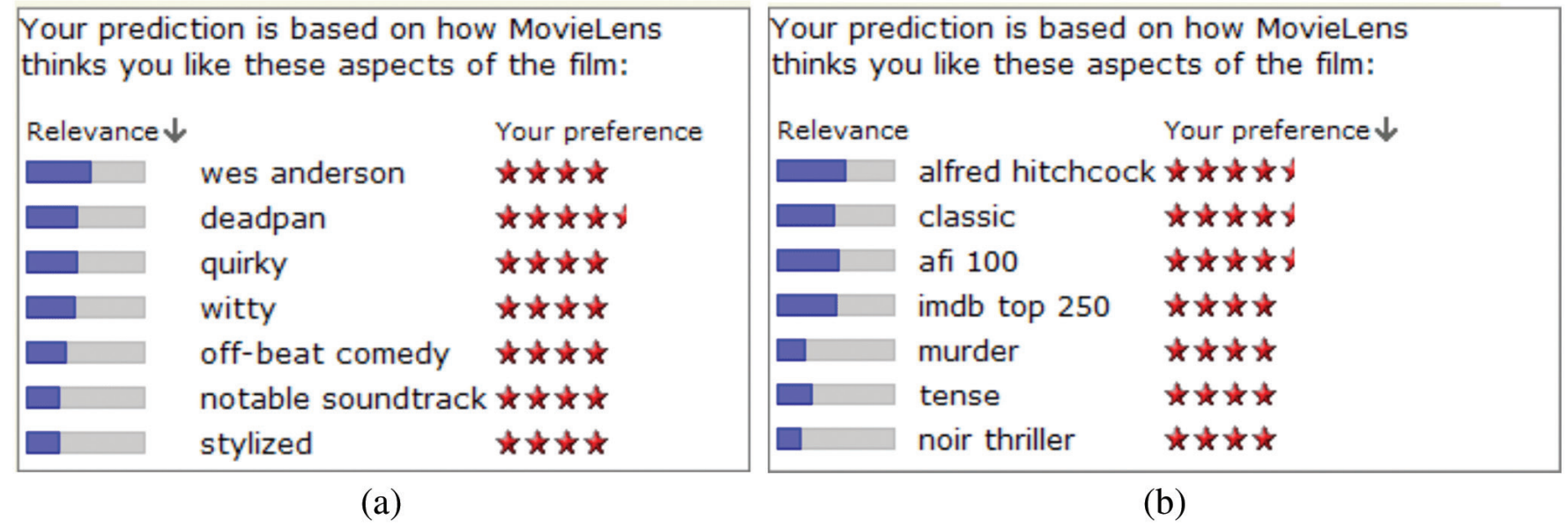

**Figure 4.1:** Some explanation interfaces for online user study in Vig *et al.* (2009). (a) RelSort interface: Shows relevance and preference, and sorts tags by relevance. (b) PrefSort interface: Shows relevance and preference, and sorts tags by preference.

User study is also used to evaluate recent machine learning approaches to explainable recommendation. For example, Wang *et al.* (2018b) recruited participants through Amazon Mechanical Turk to evaluate with a diverse population of users. The study is based on the review data in Amazon and Yelp datasets. For each participant, the authors randomly selected a user from the dataset, and showed this user's reviews to the participant, so that the participant can get familiar with the user. Participants are then asked to infer the user's preference based on these reviews. Then they will be asked to evaluate the recommendations and explanations provided to the user by answering several survey questions from this user's perspective. Except for textual explanations, user study is also used to evaluate visual explanations, for example, Chen *et al.* (2019b) generated visual explanations by highlighting image region-of-interest to users, and leveraged Amazon MTurk to hire freelancers to label the ground-truth images for evaluation.

Besides large-scale online workers, we may also conduct user studies with relative small-scale volunteers, paid subjects, or manually labeling the explanations. For example, Wang and Benbasat (2007) adopted a survey-based user study approach to investigate the trust and understandability of content-based explanations. They examined the effects of three explanation types – how, why, and trade-off explanations – on consumers' trusting beliefs in competence, benevolence, and integrity. The authors built a recommendation system experimental platform, and results confirmed the critical role of explanation in enhancing consumers' initial trusting beliefs. Ren *et al.* (2017) took a random sample of 100 recommendations, and manually evaluated their explanations regarding the accuracy of sentiment labels, which helped to verify that the proposed viewpoint-based explanations are more informative than topic labels in prior work.

## 4.2 Online Evaluation

Another approach to evaluating explainable recommendation is through online experiments. There could be several different perspectives to consider, including persuasiveness, effectiveness, efficiency, and satisfaction of the explanations.

Due to the limited type of information that one can collect in online systems, it is usually easier to evaluate the persuasiveness of the explanations, i.e., to see if the explanations can help to make users accept the recommendations. For example, Zhang *et al.* (2014a) conducted online experiments focusing on how the explanations affect user acceptance. The authors conducted A/B-tests based on a commercial web browser, which has more than 100 million users, with 26% monthly active users. The experiments recommend compatible phones when a user is browsing mobile phones in an online shopping website, as shown in Figure 4.2. To evaluate the explanation persuasiveness, three user groups are designed, including an experimental group that receives the testing explanations, a comparison group that receives the baseline "People also viewed" explanations, and a control group that receives no explanation. The click-through rate of each group is calculated to evaluate the effect of providing personalized explanations. Besides, they also calculated the percentage of recommendations added to cart by users to evaluate the conversation rate, and the percentage of agreements to evaluate the explanation effectiveness, where a recommendation (or a dis-recommendation) is considered as an agreement if it was (or was not) added to cart, respectively.

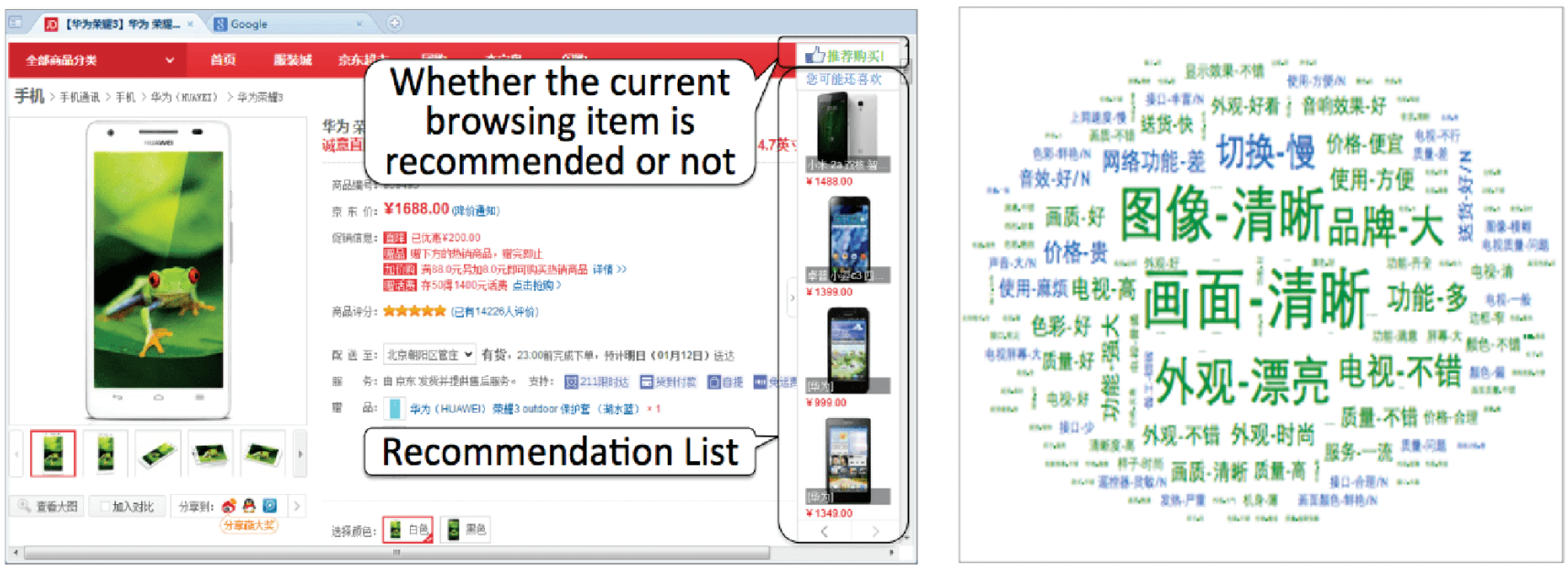

**Figure 4.2:** Top-4 recommended items are presented by the browser at the right-hand side when the user is browsing an online product. A feature-opinion word cloud is also displayed to assist explanations. For example, the biggest pair in the right figure means "PictureClarity-High". Explanations are displayed only when the user hovers the mouse on a recommended item. In this way, the system knows that the user indeed examines the explanation.

We should note that the evaluation measures in online scenarios could vary depending on the availability of the resources in the testing environment. For example, one may evaluate based on click-through-rate (CTR) when user click information is available, or calculate the purchase rate if user purchase actions can be tracked, or even calculate the gross profit if product price information is available.

Online evaluation and user study may be concurrently performed, i.e., sometimes user study can be performed in online platforms, however, they also have significant differences regardless the operating platform. User study usually asks participants to complete certain questions or tasks under given experimental instructions, as a result, participants usually know that they are being investigated. Online evaluation such as A/B testing, on the other hand, usually exposures selected users in the experimental environment unconsciously, and then collects user behaviors therein to study and compare the effectiveness of different strategies. Both approaches have pros and cons, e.g., due to the flexibility of questions to ask, user study enables researchers to design complex surveys to analyze the problem at hand in depth, while in online evaluation, researchers have to make comparisons based on a few certain types of user behaviors. On the other hand, the fact that participants know their being experimented in user study may bring cognitive bias to the results, as a result, a lot of research efforts have been made to reduce the bias in user studies (Arnott, 2006).

## 4.3 Offline Evaluation

In general, there are two approaches to evaluating recommendation explanations offline. One is to evaluate the percentage of recommendations that can be explained by the explanation model, regardless of the explanation quality; and the second approach is to evaluate the explanation quality directly. However, we have to note that more offline evaluation measures/protocols are yet to be proposed for comprehensive explainability evaluation.

For the first approach, Abdollahi and Nasraoui (2017) adopted mean explainability precision (MEP) and mean explainability recall (MER). More specifically, explainability precision (EP) is defined as the proportion of explainable items in the top-$n$ recommendation list, relative to the total number of recommended (top-$n$) items for each user. Explainability recall (ER), on the other hand, is the proportion of explainable items in the top-$n$ recommendation list, relative to the total number of explainable items for a given user. Finally, mean explainability precision (MEP) and mean explainability recall (MER) are EP and ER averaged across all testing users, respectively. Peake and Wang (2018) further generalized the idea and proposed model *Fidelity* as a measure to evaluate explainable recommendation algorithms, which is defined as the percentage of explainable items in the recommended items:

$$\text{Model Fidelity} = \frac{|\text{explainable items} \cap \text{recommended items}|}{|\text{recommended items}|}$$

For the second approach, evaluating the quality of the explanations usually depends on the type of explanations. One commonly used explanation type is a piece of explanation sentence. In this case, offline evaluation can be conducted with text-based measures. For example, in many online review websites (such as e-commerce), we can consider a user's true review for an item as the ground-truth explanation for the user to purchase the item. If our explanation is a textual sentence, we can take frequently used text generation measures for evaluation, such as BLEU score (bilingual evaluation understudy, Papineni *et al.*, 2002) and ROUGE score (recall-oriented understudy for gisting evaluation, Lin, 2004). The explanation quality can also be evaluated in terms of readability measures, such as Gunning Fog Index (Gunning, 1952), Flesch Reading Ease (Flesch, 1948), Flesch Kincaid Grade Level (Kincaid *et al.*, 1975), Automated Readability Index (Senter and Smith, 1967), and Smog Index (Mc Laughlin, 1969).

Overall, regardless of the explanation style (text or image or others), offline explanation quality evaluation would be easy if we have (small scale) ground-truth explanations. In this way, we can evaluate how well the generated explanations match with the ground-truth, in terms of precision, recall, and their variants.

## 4.4 Qualitative Evaluation by Case Study

Case study as the qualitative analysis is also frequently used for explainable recommendation research. Providing case studies can help to understand the intuition behind the explainable recommendation model and the effectiveness of explanations. Providing case studies as qualitative analysis also helps to understand when the proposed approach works and when it does not work.

For example, Chen *et al.* (2018c) provided case study to explain the sequential recommendations, as shown in Figure 4.3. Through case studies, the authors found that many sequential recommendations can be explained by "one-to-multiple" or "one-to-one" user behavior patterns. "One-to-multiple" means that a series of subsequent purchases are triggered by the same item, while "one-to-one" means that each of the subsequent purchases is triggered by its preceding item. These explanations can help users to understand why an item is recommended and how the recommended items match their already purchased items.

Hou *et al.* (2018) adopted case studies to analyze the user preference, item quality, and explainability of the hotel recommendations. The authors first proposed a metric called Satisfaction Degree on Aspects (SDA) to measure the user satisfaction on item aspects, and then conducted case studies to show how the model explains the recommendations, as shown in Figure 4.4. In this example, item 1 is recommended instead of item 2 for the target user. By examining the user preference and item quality, this recommendation is explained by the fact that item 1 satisfies user preferences over most aspects.

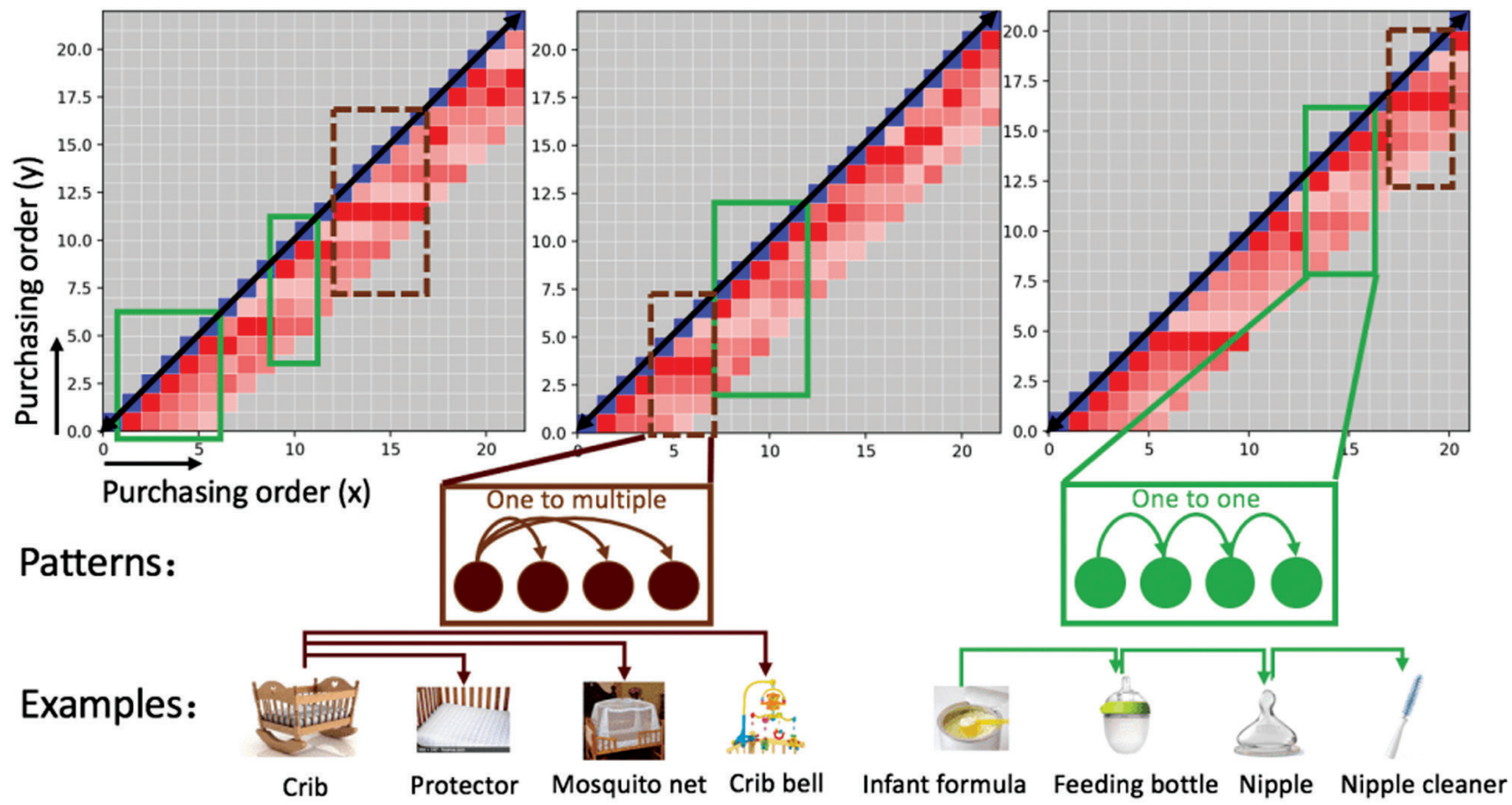

**Figure 4.3:** Case study of e-commerce user behaviors for explainable sequential recommendations. Recommendations could be explained by "one-to-multiple" behaviors, e.g., a baby crib leads to a mattress protector, a mosquito net, and a bed bell; recommendations may also be explained by "one-to-one" behaviors, e.g., an infant formula leads to a feeding bottle, nipples, and a nipple cleaner step by step.

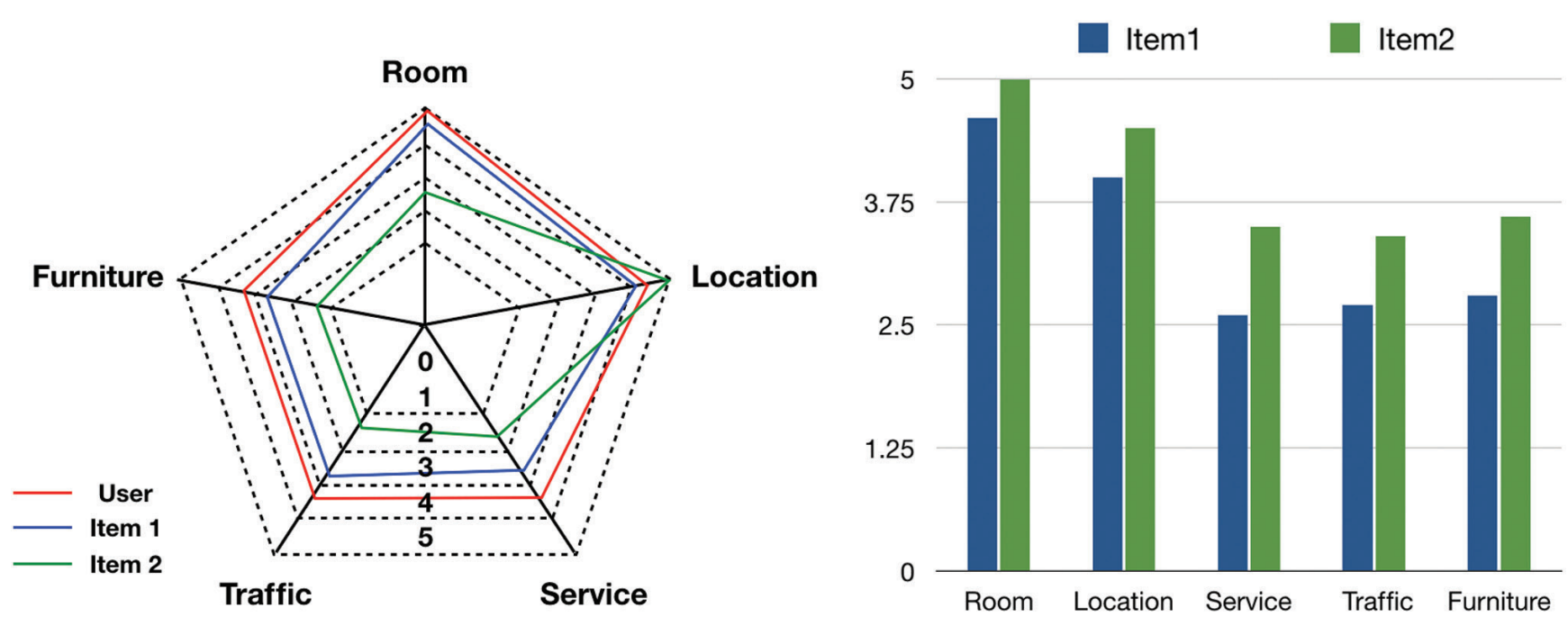

**Figure 4.4:** Case study of explainable hotel recommendations for a target user. For each item, the algorithm calculates its quality on each aspect, and for each user, the algorithm calculates user preference on each aspect. The system then draws a radar chart and a bar chart of the user preferences and item qualities for explanation.

## 4.5 Summary

In this section, we introduced the evaluation methods for explainable recommendation research. A desirable explainable recommendation model would not only be able to provide high-quality recommendations but also high-quality explanations. As a result, an explainable recommendation model would better be evaluated in terms of both perspectives.

In this section, we first presented frequently used evaluation methods for recommendation systems, including both online and offline approaches. We further introduced methods particularly for explanation evaluation, including both quantitative and qualitative methods. More specifically, quantitative methods include online, offline, and user study approaches, while qualitative evaluation is usually implemented by case studies over the generated explanations.

# 5

# Explainable Recommendation in Different Applications

The research and application of explainable recommendation methods span across many different scenarios, such as explainable e-commerce recommendation, explainable social recommendation, and explainable multimedia recommendation.

In this section, we provide a review of explainable recommendation methods in different applications. Most of the research papers in this section have already been introduced in previous sections. Instead, we organize them based on their application scenario to help readers better understand the current scope of explainable recommendation research and how it helps in different applications.

## 5.1 Explainable E-commerce Recommendation

Product recommendation in e-commerce is one of the most widely adopted scenarios for explainable recommendations. It has been a standard test setting for explainable recommendation research.

As an example of this scenario, Zhang *et al.* (2014a) proposed explainable recommendation based on the explicit factor model, and conducted online experiments to evaluate the explainable recommendations based on a commercial e-commerce website (JD.com). Later, many explainable recommendation models are proposed for e-commerce recommendation. For instance, He *et al.* (2015) introduced a tripartite graph ranking algorithm for explainable recommendation of electronics products; Chen *et al.* (2016) proposed a learning to rank approach to cross-category explainable recommendation of the products; Seo *et al.* (2017) and Wu *et al.* (2019) conducted explainable recommendation for multiple product categories in Amazon, and highlighted important words in user reviews based on attention mechanism; Heckel *et al.* (2017) adopted overlapping co-clustering to provide scalable and interpretable product recommendations; Chen *et al.* (2019b) proposed a visually explainable recommendation model to provide visual explanations for fashion products; Hou *et al.* (2018) used product aspects to conduct explainable video game recommendation in Amazon; Chen *et al.* (2018a) leveraged neural attention regression based on reviews to conduct rating prediction on three Amazon product categories; Chen *et al.* (2018c) adopted memory networks to provide explainable sequential recommendations in Amazon; Wang *et al.* (2018b) leveraged multi-task learning with tensor factorization to learn textual explanations for Amazon product recommendation; By incorporating explicit queries, explainable recommendation can also be extended to explainable product search in e-commerce systems (Ai *et al.*, 2019).

Explainable recommendations are essential for e-commerce systems, not only because it helps to increase the persuasiveness of the recommendations, but also because it helps users to make efficient and informed decisions (Schafer *et al.*, 2001). Since more and more consumer purchases are made in the online economy, it becomes essential for e-commerce systems to be socially responsible by achieving commercial profits and benefiting consumers with the right decisions simultaneously. Explainable recommendation – as a way to help users understand why or why not a product is the right choice – is an important technical approach to achieving the ultimate goal of socially responsible recommendations.

## 5.2 Explainable Point-of-Interest Recommendation

Point-of-Interest (POI) recommendation – or location recommendation in a broader sense – tries to recommend users with potential locations of interest, such as hotels, restaurants, or museums. Explainable POI recommendation gained considerable interest in recent years. Most of the research is based on datasets from POI review websites, such as Yelp[1] and TripAdvisor.[2]

By providing appropriate explanations in POI recommendation systems, it helps users to save time and minimize the opportunity cost of making wrong decisions, because traveling from one place to another usually means extensive efforts in time and money. Besides, providing explanations in travel planning applications (such as TripAdvisor) helps users better understand the relationship between different places, which could help users to plan better trip routes in advance.

In terms of explainable POI recommendation research, Wu and Ester (2015) conducted Yelp restaurant recommendation and TripAdvisor hotel recommendation. The authors proposed a probabilistic model combining aspect-based opinion mining and collaborative filtering to provide explainable recommendations, and the recommended locations are explained by a word cloud of location aspects. Bauman *et al.* (2017) developed models to extract the most valuable aspects from reviews for the restaurant, hotel, and beauty&spa recommendations on Yelp. Seo *et al.* (2017) also conducted explainable restaurant recommendations on Yelp. The authors proposed interpretable convolutional neural networks to highlight informative review words as explanations. Zhao *et al.* (2015) conducted POI recommendation based on Yelp data in the Phoenix city and Singapore, respectively, and the authors proposed a joint sentiment-aspect-region modeling approach to generating recommendations. Wang *et al.* (2018c) proposed a tree-enhanced embedding model for explainable tourist and restaurant recommendation based on TripAdvisor data in London and New York. Baral *et al.* (2018) proposed a dense subgraph extraction model based on user-aspect bipartite graphs for explainable location recommendation in location-based social networks. The results have shown that providing appropriate explanations increases the user acceptance on the location recommendations.

---

[1] http://www.yelp.com

[2] http://www.tripadvisor.com

## 5.3 Explainable Social Recommendation

Explainable recommendations also apply to social environments. Prominent examples include friend recommendations, news feeding recommendations, and the recommendation of blogs, news, music, travel plans, web pages, images, or tags in social environments.

Explainability of the social recommender systems is vitally important to the users' trustworthiness in the recommendations, and trustworthiness is fundamental to maintain the sustainability of social networks (Sherchan *et al.*, 2013). For example, by providing the overlapping friends as explanations for friend recommendations on Facebook, it helps users to understand why an unknown person is related to them and why the recommended friend would be trusted. By telling the user which of his or her friends have twitted a piece of news as explanations in Twitter, it helps the user to understand why the recommended news could be important for herself. It also helps users to quickly identify useful information to save time in the era of information overload.

Explainable recommendations in social environments are also important to the credibility of news recommendations (Bountouridis *et al.*, 2018). Since any individual can post and re-post news articles in the social environment, the system may be exploited to spread fake news or make unjustified influences on our society. By explaining the credibility of news articles based on cross-referencing (Bountouridis *et al.*, 2018), we can help users to identify credible vs. fake information in social environments, which is critical for national security.

In terms of the research on social explainable recommendation, Ren *et al.* (2017) proposed a social collaborative viewpoint regression model for rating prediction based on user opinions and social relations. The social relations not only help to improve the recommendation performance, but also the explainability of the recommendations. Quijano-Sanchez *et al.* (2017) developed a social explanation system applied to group recommendations. It integrates explanations about the group recommendations and explanations about the group's social reality, which gives better perceptions of the group recommendations. Tsai and Brusilovsky (2018) studied how to design explanation interfaces for casual (non-expert) users to achieve different explanatory goals. In particular, the authors conducted an international online survey of a social recommender system – based on 14 active users and across 13 countries – to capture user feedback and frame it in terms of design principles of explainable social recommender systems. The research results have shown that explanations in social networks help to benefit social network users in terms of transparency, scrutability, trust, persuasiveness, effectiveness, efficiency and satisfaction.

## 5.4 Explainable Multimedia Recommendation

Explainable multimedia recommendation broadly includes the explainable recommendation of books (Wang *et al.*, 2018a), news/articles (Kraus, 2016), music (Celma, 2010; Zhao *et al.*, 2019a), movies (Herlocker *et al.*, 2000; Nanou *et al.*, 2010; Tintarev and Masthoff, 2008), or videos (Toderici *et al.*, 2010). A prominent example is the YouTube recommendation engine. Providing explanations for multimedia recommendations can help users to make informed decisions more efficiently, which further helps to save time by reducing unnecessary data-intensive media transmissions through the network. We review the related work on explainable multimedia recommendation in this section.

The MovieLens dataset[3] is one of the most frequently used datasets for movie recommendation (Herlocker *et al.*, 2000). Based on this dataset, Abdollahi and Nasraoui (2016) proposed explainable matrix factorization by learning the rating distribution of the active user's neighborhood. In Abdollahi and Nasraoui (2017), the authors further extended the idea to the explainability of constrained matrix factorization. Chang *et al.* (2016) adopted crowd-sourcing to generate crowd-based natural language explanations for movie recommendations in MovieLens. Lee and Jung (2018) provided story-based explanations for movie recommendation systems, achieved by a multi-aspect explanation and narrative analysis method.

Based on a knowledge-base of the movies, such as the genre, type, actor, and director, recent research has been trying to provide knowledge-aware explainable recommendations. For example, Catherine *et al.*

[3] https://grouplens.org/datasets/movielens/

(2017) proposed explainable entity-based recommendation with knowledge graphs for movie recommendation. It provides explanations by reasoning over the knowledge graph entities about the movies. Wang *et al.* (2018a) proposed the ripple network model to propagate user preferences on knowledge graphs for recommendations. It provides explanations based on knowledge paths from the user to the recommended movie, book, or news.

Zhao *et al.* (2019a) studied personalized reason generation for explainable song recommendation in conversational agents (e.g., Microsoft XiaoIce[4]). The authors proposed a solution that generates a natural language explanation of the reason for recommending a song to a particular user, and showed that the generated explanations significantly outperform manually selected reasons in terms of click-through rate in large-scale online environments. Davidson *et al.* (2010) introduced the YouTube Video Recommendation System, which leveraged association rule mining to find the related videos as explanations of the recommended video. Online media frequently provide news article recommendations to users. Recently, such recommendations have been integrated into independent news feeding applications on the phone, such as the Apple News. Kraus (2016) studied how news feeds can be explained based on political topics.

## 5.5 Other Explainable Recommendation Applications

Explainable recommendation is also essential to many other applications, such as academic recommendation, citation recommendation, legal recommendation, and healthcare recommendation. Though direct explainable recommendation work on these topics is still limited, researchers have begun to consider the explainability issues within these systems. For example, Gao *et al.* (2017) studied the explainability of text classification in online healthcare forums, where each sentence is classified into three types: medication, symptom, or background. An interpretation method is developed, which explicitly extracts the decision rules to gain insights about the useful information in texts.

[4] http://www.msxiaoice.com/

Liu *et al.* (2018) further studied interpretable outlier detection for health monitoring. In healthcare practice, it is usually important for doctors and patients to understand why data-driven systems recommend a certain treatment, and thus, it is important to study the explainability of healthcare recommendation systems. It is also worth noting that explainability perspectives have also been integrated into other intelligent systems beyond recommendation, such as explainable search (Singh and Anand, 2019), question answering (Zhao *et al.*, 2019b), and credibility analysis of news articles (Bountouridis *et al.*, 2018).

## 5.6 Summary

In this section, we introduced several applications of explainable recommendation to help readers understand how the key idea of explainability works in different recommendation scenarios. In particular, we introduced explainable e-commerce, POI, social, and multimedia recommendations. We also briefly touched some new explainable recommendation tasks such as explainable academic, educational, and healthcare recommendations, which have been attracting increasing attention recently.

It is also beneficial to discuss the potential limitations of explainable recommendation. Although explanations can be helpful to many recommendation scenarios, there could exist scenarios where explanations are not needed or could even hurt. These include time-critical cases where decisions should be made in real-time, and users are not expected to spend time evaluating the decisions. For example, while driving on highways, users may want to know the correct exit directly without spending time listening to the explanations. Even more critical scenarios include emergency medical decisions or battlefield decisions, where spending time for evaluation may not be permitted. Depending on the scenario, explainable recommendation systems may need to avoid providing too much explanation, and avoid repeated explanations, explaining the obvious, or explaining in too many details, which may hurt rather than improve the user experience. The system also needs to be especially careful not to provide obviously misleading or even wrong explanations, because people may lose confidence in algorithms after seeing them trying to mistakenly convince others (Dietvorst *et al.*, 2015).

# 6

# Open Directions and New Perspectives

We discuss some open research directions and new research perspectives of explainable recommendation in this section. We make discussion on four broad perspectives: methodology, evaluation, cognitive foundation, and broader impacts.

## 6.1 Methods and New Applications

### 6.1.1 Explainable Deep Learning for Recommendation

The research community has been developing explainable deep learning models for explainable recommendations. Current approaches focus on designing deep models to generate explanations accompanying the recommendation results. The explanations could come from attention weights over texts, images, or video frames. However, the research of explainable deep learning is still in its initial stage, and there is still much to explore in the future (Gunning, 2017).

Except for designing deep models for explainable recommendations, the explainability of the deep model itself also needs further research. In most cases, the recommendation and explanation models are still black boxes, and we do not fully understand how an item is recommended out of other alternatives. This is mostly because the hidden layers in most deep neural networks do not possess intuitive meanings. As a result, an important task is to make the deep models explainable for recommendations. This will benefit not only the personalized recommendation research, but also many other research areas such as computer vision and natural language processing, as well as their application in healthcare, education, chatbots, and autonomous systems, etc.

Recent advances in machine learning have shed light on this problem, for example, Koh and Liang (2017) provided a framework to analyze deep neural networks based on influence analyses, while Pei *et al.* (2017) proposed a white-box testing mechanism to help understand the nature of deep learning systems. Regarding explainable recommendation, this will help us to understand what are the meanings of each latent component in a neural network, and how they interact with each other to generate the final results.

### 6.1.2 Knowledge-enhanced Explainable Recommendation

Most of the explainable recommendation research is based on unstructured data, such as texts or images. However, if the recommendation system possesses specific knowledge about the recommendation domain, it will help to generate more tailored recommendations and explanations. For example, with the knowledge graph about movies, actors, and directors, the system can explain to users precisely that "a movie is recommended because he has watched many movies starred by an actor". Such explanations usually have high fidelity scores. Previous work based on this idea dates back to the content-based recommendation, which is effective, but lacks serendipity and requires extensive manual efforts to match the user interests with the content profiles.

With the fast progress of (knowledge) graph embedding techniques, it has been possible for us to integrate the learning of graph embeddings and recommendation models for explainable recommendation, so that the system can make recommendations with specific domain knowledge, and tell the user why such items are recommended based on knowledge reasoning, similar to what humans do when asked to make recommendations. It will also help to construct conversational recommendation systems, which communicate with users to provide explainable recommendations based on knowledge. Moreover, in a more general sense, this represents an important future direction for intelligent systems research, i.e., to integrate rational and empirical approaches for agent modeling.

### 6.1.3 Multi-Modality and Heterogenous Information Modeling

Modern information retrieval and recommendation systems work on many heterogeneous multi-modal information sources. For example, web search engines have access to documents, images, videos, and audios as candidate search results; e-commerce recommendation system works on user numerical ratings, textual reviews, product images, demographic information and others for user personalization and recommendation; social networks leverage user social relations and contextual information such as time and location for search and recommendation.

Current systems mostly leverage heterogeneous information sources to improve search and recommendation performance, while many research efforts are needed to use heterogeneous information for explainability. These include a wide range of research tasks such as multi-modal explanations by aligning two or more different information sources, transfer learning over heterogeneous information sources for explainable recommendations, cross-domain explanation in information retrieval and recommendation systems, and how the different information modalities influence user receptiveness on the explanations.

### 6.1.4 Context-aware Explanations

User preferences or item profiles may change along with context information such as time and location, and thus personalized recommendations could be context-aware (Adomavicius and Tuzhilin, 2011). The same idea applies to explainable recommendations. Because user preferences may change over context, the explanations could also be context-aware, so that recommendations can be explained in the most appropriate way. Most of the current explainable recommendation models are static, i.e., users are profiled based on a training dataset, and explanations are generated accordingly, while context-aware explainable recommendation needs extensive exploration in the future.

### 6.1.5 Aggregation of Different Explanations

Different explainable recommendation models may generate different explanations, and the explanations may highly depend on the specific model. As a result, we usually have to design different explainable models to generate different explanations for different purposes. On one hand, researchers have shown that providing diversified explanations is beneficial to user satisfaction in recommender systems (Tsukuda and Goto, 2019). While on the other hand, different explanations may not be logically consistent in explaining one item, and according to cognitive science research, having a complete set of explanations may not be what we need in many cases (Miller, 2019). When the system generates many candidate explanations for a search or recommendation result, a significant challenge is how to select the best combination of the explanations to display, and how to aggregate different explanations into a logically consistent unified explanation. Solving this problem may require extensive efforts to integrate statistical and logical reasoning approaches to machine learning, so that the decision-making system is equipped with the ability of logical inference to explain the results.

### 6.1.6 Explainable Recommendation as Reasoning

Early approaches to AI, such as Logical/symbolic AI – also known as the Good Old-Fashioned Artificial Intelligence (GOFAI) – were highly transparent and explainable, but they are less effective in generalization and robustness to noise. Later approaches such as machine learning and recent deep learning are more effective in learning patterns from data for prediction, but they are less transparent due to the model complexity and latent computation.

As a representative branch of AI research, the advancement of collaborative filtering for recommendation followed a similar path. Early approaches to CF adopted straightforward yet transparent methods, such as user-based (Resnick *et al.*, 1994) or item-based (Sarwar *et al.*, 2001) methods, which find similar users or items first, and then calculate the weighted average ratings for prediction. Later approaches to CF, however, more and more advanced to less transparent “latent” machine learning approaches, beginning from shallow latent factor models such as matrix factorization to deep learning approaches. Though effective in ranking and rating prediction, the key philosophy of these approaches is to learn user/item representations and similarity networks from data for better user-item matching, which makes it difficult to understand how the prediction network generated the output results, because the results are usually produced by latent similarity matching instead of an explicit reasoning procedure. To solve the problem, we need to integrate the advantages of symbolic reasoning and machine learning – such as causal reasoning and neural logic reasoning – so as to advance collaborative filtering to collaborative reasoning for both better recommendations and better explainability (Chen *et al.*, 2020; Shi *et al.*, 2019).

### 6.1.7 NLP and Explainable Recommendation

Most explainable recommendation models are designed to generate some predefined types of explanations, e.g., based on sentence templates, specific association rules, or word clouds. A more natural explanation form could be free-text explanations based on natural language.

There has been some work trying to generate natural language explanations. The basic idea is to train sentence generation models based on user reviews and generate "review-like" sentences as explanations, such as Li *et al.* (2017), Costa *et al.* (2018), Chen *et al.* (2019d), and Ni *et al.* (2019). The research of generating natural language explanation is still in its early stage, and much needs to be done so that machines can explain themselves using natural language. For example, not all of the review contents are of explanation purpose, and it is challenging to cope with various noise for generating explanations. Since explanations should be personalized, it is important to adapt pre-trained language models such as BERT (Devlin *et al.*, 2018) to personalized per-training models. Explanations should also be generated beyond textual reviews, e.g., we can integrate visual images, knowledge graphs, sentiments, and other external information to generate more informed natural language explanations, such as explanation with specific sentiment orientations. Natural language explanations are also crucial for explainable conversational systems, which we will discuss in the following subsections.

### 6.1.8 Answering the "Why" in Conversations

The research of recommendation system has extended itself to multiple perspectives, including *what* to recommend (user/item profiling), *when* to recommend (time-aware), *where* to recommend (location-based), and *who* to recommend (social recommendation). Beyond these, explainable recommendation aims at answering the question of *why* to recommend, which attempts to solve the problem of users' inherent curiosity about why a recommended item is suitable for him or her. Demonstrating why an item is recommended not only helps users to understand the rationality of the recommendations, but also helps to improve the system efficiency, transparency, and trustworthiness.

Based on different application scenarios, users can receive recommendation explanations either passively or actively. In conventional web-based systems such as online e-commerce, the explanations can be displayed together with the recommended item, so that the users passively receive the explanations for each recommendation. In the emerging environment of conversational recommendation based on smart agent devices, users can ask "why-related" questions to actively seek for explanations when a recommendation is not intuitive. In this case, explainable recommendations will significantly increase the scope of queries that intelligent systems can process.

## 6.2 Evaluation and User Behavior Analysis

### 6.2.1 Evaluation of Explainable Recommendations

Evaluation of explainable recommendation systems remains a significant problem. For recommendation performance, explainable recommendations can be easily evaluated based on traditional rating prediction or top-n ranking measures, while for explanation performance, a reliable protocol is to test explainable vs. non-explainable recommendations based on real-world user studies, such as A/B testing in practical systems, and evaluation with paid subjects or online workers on Mechanical Turk. However, there still lacks a usable offline *explainability* measure to evaluate the explanation performance.

Evaluation of the explanations is related to both user perspective and algorithm perspective. On user perspective, the evaluation should reflect how the explanations influence users, in terms of, e.g., persuasiveness, effectiveness, efficiency, transparency, trustworthiness, and user satisfaction. On algorithm perspective, the evaluation should reflect to what degree the explanations reveal the real mechanism that generated the recommendations (sometimes called explanation fidelity). The evaluation may also be related to the form of explanations, e.g., visual and textual explanations could be evaluated in different ways. Developing reliable and readily usable evaluation measures for different evaluation perspectives will save many efforts for offline evaluation of explainable recommendation systems.

### 6.2.2 User Behavior Perspectives

Though early research explored a lot about how users interact with explanations (Herlocker *et al.*, 2000), many recent research on explainable recommendation is mostly model-driven, which designs new explainable models to generate recommendations and explanations. However, since recommender systems are inherently human-computer interaction systems, it is also essential to study explainable recommendations from user behavior perspectives, where the "user" could either be recommender system designers or normal users of the system. There exist a broad scope of problems to explore, including but not limited to how users interact with explanations, and the user receptiveness on different types of explanations. Research on user behavior perspectives for explainable recommendation will also benefit the evaluation of explainable recommendation systems.

## 6.3 Explanation for Broader Impacts

Existing explainable recommendations mostly focus on generating explanations to *persuade* users to accept the explanations (Nanou *et al.*, 2010). However, explanations could have broader impacts beyond persuasiveness. It is worthwhile to explore how explanations can help to improve the trustworthiness (Cramer *et al.*, 2008), efficiency (Tintarev and Masthoff, 2011), diversity (Yu *et al.*, 2009), satisfaction (Bilgic and Mooney, 2005), and scrutability of the system (Balog *et al.*, 2019; Knijnenburg *et al.*, 2012). For example, by letting the user know why not to buy a certain product, the system can help to save time for the users and to win the user's trust in the system (Zhang *et al.*, 2014a).

Another important problem is the relationship between explainability and fairness. Though researchers have shown that transparency helps to increase the fairness in economic (Cowgill and Tucker, 2019), political (Fine Licht, 2014) and legal (Burke and Leben, 2007) systems, we are yet to see if and how transparency and fairness relate with each other in information systems. One example is that if the system inevitably has to output unfair rankings for a user, explaining to the user why this happens could help to gain user's understanding.

## 6.4 Cognitive Science Foundations

There are two research philosophies towards explainable recommendation – and more broadly explainable decision making or explainable AI (Lipton, 2018). One is to design transparent/interpretable models directly for decision making, and such models usually naturally output the decision explanations. Another philosophy is that we only focus on the explainability of the recommendation results. In this way, we still treat the recommendation model as a blackbox, and instead develop separate models to explain the results produced by this blackbox. Most of the post-hoc/model-agnostic approaches adopt this philosophy.

It is interesting to see that both research philosophies have their roots in human cognitive science (Miller, 2019). Sometimes, human-beings make decisions based on careful reasoning, and they can clearly explain how a decision is made by showing the reasoning process step by step. Other times, people make intuition decisions first and then "seek" an explanation for the decision, which belongs to the post-hoc explanation approach. It is challenging to decide which research philosophy towards explainable recommendation – and explainable AI in a broader sense – is the correct approach (or both). Answering this question requires significant breakthroughs in human cognitive science as well as our understanding of how the human brain works.

# 7

# Conclusions

Early recommendation models – such as user/item-based collaborative filtering – were very transparent and explainable, and the lack of transparency of best match ranking models has been known and studied as an important downside from the start. Recent advances on more complex models – such as latent factor models or deep representation learning – helped to improve the search and recommendation performance, but they further brought about the difficulty of transparency and explainability.

The lack of explainability mainly exists in terms of two perspectives: 1) the outputs of the recommendation system (i.e., recommendation results) are hardly explainable to system users, and 2) the mechanism of the recommendation model (i.e., recommendation algorithm) is hardly explainable to system designers. This lack of explainability for recommendation algorithms leads to many problems: without letting the users know why specific results are provided, the system may be less effective in persuading the users to accept the results, which may further decrease the system's trustworthiness. More importantly, many recommendation systems nowadays are not only useful for information seeking – by providing supportive information and evidence, they are also crucial for complicated decision making. For example, medical workers may need comprehensive healthcare document recommendations or retrieval to make medical diagnoses. In these decision making tasks, the explainability of the results and systems is extremely important, so that system users can understand why a particular result is provided, and how to leverage the result to take actions.

Recently, deep neural models have been used in many information retrieval and recommendation systems. The complexity and inexplainability of many neural models have further highlighted the importance of explainable recommendation and search, and there is a wide range of research topics for the community to address in the coming years.

In this survey, we provided a brief history of explainable recommendation research ever since the early stage of recommendation systems, towards the very recent research achievements. We introduced some different types of explanations, including user/item-based, content-based, textual, visual, and social explanations. We also introduced different explainable recommendation models, including MF-based, topic-based, graph-based, deep learning-based, knowledge-based, mining-based, and post-hoc explainable recommendation models. We further summarized representative evaluation methods for explainable recommendations, as well as different explainable recommendation applications, including but not limited to, explainable e-commerce, POI, social, and multimedia recommendations. As an outlook to the future, we summarized several possible new perspectives on the explainable recommendation research. We expect that knowledge-graphs, deep learning, natural language processing, user behavior analysis, model aggregation, logical reasoning, conversational systems, and cognitive foundations to help advance the development of explainable recommendation.

In a broader sense, researchers in the broader AI community have also realized the importance of Explainable AI, which aims to address a wide range of AI explainability problems in deep learning, computer vision, autonomous driving systems, and natural language processing tasks. As an essential branch of AI research, this highlights the importance of the IR/RecSys community to address the explainability issues of various search and recommendation systems.

# Acknowledgements

We sincerely thank the reviewers for providing the valuable reviews and constructive suggestions.

The work is partially supported by National Science Foundation (IIS-1910154). Any opinions, findings and conclusions expressed in this material are those of the authors and do not necessarily reflect those of the sponsors.